%% file: main.tex
\documentclass[sigconf]{acmart}
\usepackage{array}
\newcolumntype{L}{@{}>{\iffalse}l<{\fi}}
\usepackage{etex}
\usepackage{wrapfig}
\usepackage{booktabs} 
\usepackage[shortcuts]{extdash}
\usepackage{mathtools}
\usepackage{mathpartir}
\usepackage{amssymb} 
\usepackage{macros}
\usepackage{stmaryrd}
\usepackage{listings}
\usepackage{bigstrut}
\usepackage{color}
\usepackage{multirow}
\usepackage{courier}
\usepackage{paralist}
\usepackage[utf8]{inputenc}
\usepackage[disable]{todonotes}   
\usepackage[boxed]{algorithm2e}
\SetCommentSty{textit}
\usepackage{arydshln}
\usepackage{multirow}
\usetikzlibrary{matrix,positioning}
\usepackage{float}
\usepackage{placeins}
\usepackage{color}
\usepackage{pgfplots}
\usepackage{enumitem}
\newcommand{\squeezeup}{\vspace{-9mm}}
\PassOptionsToPackage{hyphens}{url}\usepackage{hyperref}

\newcommand{\myparagraph}[1]{\smallskip \noindent{\bf #1}}
\newcommand{\etal}{\textit{et al. }}
\setcopyright{none}

\newcommand{\Ra}{\Rightarrow}
\newcommand{\mydef}{\triangleq}

\begin{document}

\title{Freeing Testers from Polluting Test Objectives}


 \author{Micha\"el Marcozzi, S\'ebastien Bardin, \\ Nikolai Kosmatov, Virgile Prevosto,\\ Lo\"ic Correnson}
 \affiliation{%
   \institution{CEA, LIST, Software Reliability Laboratory}
   \city{Gif-sur-Yvette}
   \state{France}
   \postcode{91191}
 }
 \email{first.second@cea.fr}
 \author{Mike Papadakis}
 \affiliation{%
   \institution{Interdisciplinary Centre for \\Security, Reliability and Trust\\ University of Luxembourg}
   \city{Luxembourg}
   \state{Luxembourg}
   \postcode{1855}
 }
 \email{michail.papadakis@uni.lu}

\renewcommand{\shortauthors}{Marcozzi et al.}

\begin{abstract}
Testing is the primary approach for detecting software defects. A major challenge faced by testers lies in crafting efficient test suites, able to detect a maximum number of bugs with manageable effort. To do so, they rely on coverage criteria, which define some precise test objectives to be covered. However, many common criteria specify a significant number of objectives that occur to be infeasible or redundant in practice, like covering dead code or semantically equal mutants. Such objectives are well-known to be harmful to the design of test suites, impacting both the efficiency and precision of testers' effort. This work introduces a sound and scalable formal technique able to prune out a significant part of the infeasible and redundant objectives produced by a large panel of white-box criteria. In a nutshell, we reduce this challenging problem to proving the validity of logical assertions in the code under test. This technique is implemented in a tool that relies on weakest-precondition calculus and SMT solving for proving the assertions. The tool is built on top of the Frama-C verification platform, which we carefully tune for our specific scalability needs. The experiments reveal that the tool can prune out up to 
27\% of test objectives in a program and scale to applications of 200K lines of code.
\end{abstract}

%
%


\maketitle

\input{introduction}
\input{motivation}
\input{background}

\input{approach}

\input{implementation}

\input{evaluation}
\input{threats}
\input{related}

\input{conclusion}


\vspace{-1mm}
\bibliographystyle{ACM-Reference-Format}
\bibliography{main} 

\input{appendices}

\end{document}

%% file: introduction.tex
\vspace{-1mm}
\section{Introduction}
\vspace{-1mm}
\label{sec:introduction}


\myparagraph{Context.} Heretofore, software testing is the primary method for detecting software defects \cite{myers,mathur,ZhuHM-97,ammann08}. It is performed by executing the programs under analysis with some inputs, in the aim of finding some unintended (defective) behavior. As the number of possible test inputs is typically enormous, testers do limit their tests in practice to a manageable but carefully crafted set of inputs, called a \textit{test suite}. To build such suites, they rely on so-called \textit{coverage criteria}, also known as adequacy or test criteria, which define the objectives of testing \cite{ZhuHM-97,ammann08}. In particular, many \textit{white-box} criteria have been proposed so far, where the test objectives are syntactic elements of the code that should be covered by running the test suite. For example, the \textit{condition coverage} criterion imposes to cover all possible outcomes of the boolean conditions appearing in program decisions, while the \textit{mutant coverage} criterion requires to differentiate the program from a set of its syntactic variants. Testers need then to design their suite of inputs to cover the corresponding test objectives, such as --- for the two aforementioned cases --- condition outcomes or mutants to kill. 

\myparagraph{Problem.} White-box testing criteria are purely syntactic and thus totally blind to the semantic of the program under analysis. As a consequence, many of the test objectives that they define may  turn out to be in practice either
\begin{enumerate}[label=(\alph*),leftmargin=*]
\item{\textit{infeasible}}: no input can satisfy them, such as dead code or equivalent mutants \cite{ammann08}, or
\item{\textit{duplicate} versions of other objectives}: satisfied by exactly the same inputs, such as semantically equal mutants \cite{Papadakis15}, or
\item{\textit{subsumed} by another objective}: satisfied by every input covering the other objective \cite{Papadakis16, KurtzAODKG16, AmmannDO14}, such as validity of a condition logically implied by another one in condition coverage.
\end{enumerate}
We refer to these three situations as \textit{polluting test objectives}, which are well-known to be harmful to the testing task \cite{Papadakis16,Papadakis15,Yates89,Weyuker93,Woodward80} for two main reasons:
\begin{itemize}[leftmargin=*]
\item While (early) software testing theory \cite{ZhuHM-97} requires all the criterion objectives to be covered, this seldom reflects the actual practice, which usually relies on test suites covering only a part of them \cite{GligoricGZSAM15}. This is due to the difficulty of generating the appropriate test inputs, but also to infeasible test objectives. Indeed, testers often cannot know whether they fail to cover them because their test suites are weak or because they are infeasible, possibly wasting a significant amount of their test budget trying to satisfy them.
\item As full objective coverage is rarely reached in practice, testers rely on the ratio of covered objectives to measure the strength of their test suites. However, the working assumption of this practice is that all objectives are of equal value. Early testing research demonstrated that this is not true \cite{Papadakis16, AmmannDO14, Chusho87, BertolinoM94}, as duplication and subsumption can make a large number of feasible test objectives redundant. Such coverable redundant objectives may
artificially deflate or inflate the coverage ratio. This skews the measurement, which may misestimate test thoroughness and fail to evaluate correctly the remaining cost to full coverage. 
\end{itemize} 

\myparagraph{Goal and Challenges.} While detecting all polluting test objectives is undecidable  \cite{Papadakis16, AmmannDO14}, our goal is to provide a technique capable to identify a significant part of them. This is a challenging task as it requires to perform complex program analyses over large sets of objectives produced by various criteria. Moreover, duplication and subsumption should be checked for each pair of objectives, a priori putting a quadratic penalty over the necessary analyses.

Although many studies have demonstrated the harmful effects of polluting objectives, to date there is no scalable technique to discard them. Most related research works (see Tables \ref{tab:comparison-existing}, \ref{tab:comparison-analysis} and Section \ref{sec:related}) focus on the equivalent mutant problem, i.e. the particular instance of infeasible test objectives for the mutant coverage criterion. These operate either in dynamic mode, i.e. mutant classification \cite{SchulerDZ09,SchulerZ13}, or in static mode, i.e.~Trivial Compiler Equivalence (TCE) \cite{Papadakis15}. Unfortunately, the dynamic methods are unsound and produce many false positives \cite{SchulerZ13, PapadakisDT14}, while the static one deals only with strong mutation and cannot detect subsumed mutants (whereas it handles duplicates in addition to infeasible ones). The LUncov technique \cite{bardin15} combines two static analyses to prune out infeasible objectives from a panel of white-box criteria in a generic way, but faces  scalability issues.

\begin{table}[htb]
{\noindent\footnotesize
\begin{tabular}{|c|c|c|c|c|c|c|}
\cline{2-7}
\multicolumn{1}{c|}{} & \multirow{2}{*}{\textbf{Sound}} & \multirow{2}{*}{\textbf{Scale}} & \multicolumn{3}{c|}{\textbf{Kind of Pollution}} & \textbf{Criterion} \\
\multicolumn{1}{c|}{} & & & \textbf{Inf.} & \textbf{Dupl.} & \textbf{Subs.} & \textbf{Genericity} \\
\hline
\textbf{Mutant class.} \cite{SchulerZ13} & $\times$ & \checkmark & \checkmark & $\times$ & $\times$ & $\times$ \\
\hline
\textbf{TCE} \cite{Papadakis15} & \checkmark & \checkmark & \checkmark & \checkmark & $\times$ & $\times$ \\
\hline
\textbf{LUncov} \cite{bardin15} & \checkmark & $\times$ & \checkmark & $\times$ & $\times$ & \checkmark \\
\hline
\hline
\textbf{LClean} (this work) & \checkmark & \checkmark & \checkmark & \checkmark & \checkmark & \checkmark \\
\hline
\end{tabular}}
\caption{Comparison with closest research techniques}
\label{tab:comparison-existing}
\end{table}\squeezeup
\begin{table}[H]
{\noindent\footnotesize
\begin{tabular}{|c|c|c|c|}
\cline{2-4}
\multicolumn{1}{c|}{} & \textbf{Analyses} & {\textbf{Scope}} & {\textbf{Acuteness}} \\
\hline
 \multirow{2}{*}{\textbf{TCE} \cite{Papadakis15}} & built-in compiler  & \multirow{2}{*}{interprocedural}  & \multirow{2}{*}{$+$}  \\
 & optimizations &   &  \\
\hline
 \multirow{2}{*}{\textbf{LUncov} \cite{bardin15}} &   value analysis and  & \multirow{2}{*}{interprocedural}  & \multirow{2}{*}{$++$} \\
& weakest-precondition & &  \\
\hline
\hline
\textbf{LClean} (this work) & weakest-precondition &   local function & $+$ \\
\hline
\end{tabular}}
\caption{Static analyses available in closest techniques}
\label{tab:comparison-analysis}
\end{table}\squeezeup


\myparagraph{Proposal.} Our intent is to provide a \textit{unified}, \textit{sound} and \textit{scalable} solution to prune out a {significant} part of \textit{polluting objectives}, including \textit{infeasible} but also \textit{duplicate} and \textit{subsumed} ones, while handling a large panel of white-box criteria in a \textit{generic} manner. To achieve this, we propose reducing the problem of finding polluting objectives for a wide range of criteria to the problem of proving the validity of logical assertions inside the code under test. These assertions can then be verified using known verification techniques. 

Our approach, called \textit{LClean}, is the first one that scales to programs composed of 200K lines of C code, while handling all types of polluting test requirements. It is also generic, in the sense that it covers most of the common code-based test criteria (described in software testing textbooks \cite{ammann08}) and it is capable of using almost any state-of-the-art verification technique. In this study, we use weakest-precondition calculus \cite{Dijkstra76} with SMT solving \cite{smt-demoura} and identify 25K polluting test objectives in fourteen C programs. 

LClean introduces two acute code analyses that focus on detection of duplicate and subsumed objectives over a limited amount of high-hit-rate pairs of objectives. This makes it possible to detect a significant number of redundant objectives while {avoiding} a quadratic penalty in computation time. The LClean tool is implemented on top of the Frama-C/LTest platform \cite{FRAMAC,Marcozzi17b}, which features \textit{strong} conceptual and technical foundations (Section \ref{sec:background}). We specifically {extend} the Frama-C module dedicated to proving code assertions to make the proposed solution scalable and robust. 

\myparagraph{Contributions.} To sum up, we make the following contributions:
\begin{itemize}
[leftmargin=*]
\item The \textit{LClean} approach: a scalable, sound and unified formal technique (Sections \ref{sec:motivation} and \ref{sec:approach}) capable to detect the three kinds of  polluting test 
 objectives (i.e. infeasible, duplicate and subsumed) for a wide panel of white-box criteria, ranging from condition coverage to variants of MCDC and weak mutation.  
\item An \textit{open-source prototype tool LClean} (Section \ref{sec:implementation}) enacting the proposed approach. It relies on an industrial-proof formal verification platform, which we tune for the specific scalability needs of LClean, yielding a  robust multi-core assertion-proving kernel.  
\item A thorough \textit{evaluation} (Sections \ref{sec:evaluation} and \ref{sec:threats}) assessing 
\begin{enumerate}[label=(\alph*)]
\item the scalability and detection power of LClean for three types of polluting objectives and four test criteria -- pruning out up to 27\% of the objectives in C files up to 200K lines,
\item the impact of using a multi-core kernel and tailored verification libraries on the required computation time (yielding a speedup of approximately 45$\times$), and
\item that, compared to the existing methods, LClean  prunes out four times more objectives than LUncov \cite{bardin15} in about half as much time,  
it can be one order of magnitude faster than  (unsound) dynamic identification of (likely) polluting objectives, 
and  it  detects half more duplicate objectives  than TCE, while being  complementary to it.   
\end{enumerate}
\end{itemize}

\myparagraph{Potential Impact.} Infeasible test objectives have been recognized as a main cost factor of the testing process \cite{Yates89,Weyuker93,Woodward80}. By pruning out a significant number of them with LClean, testers could reinvest the gained cost in targeting full coverage of the remaining objectives. This would make testing more efficient, as most faults are found within high levels of coverage \cite{Frankl98}. Pruning out infeasible test objectives could also make the most meticulous testing criteria less expensive and thus more acceptable in industry \cite{Papadakis15}. Furthermore, getting rid of redundant objectives should provide testers with more accurate quality evaluations of their test suites and also result in sounder comparisons of test generation techniques \cite{Papadakis16}. 

%% file: motivation.tex
\section{Motivating Example}
\label{sec:motivation}


\input FigToyTrityp.tex

Figure ~\ref{fig:ToyTrityp} shows a toy C program inspired by the classic triangle 
example \cite{Myers_Art_2004}.
Given three integers $x$, $y$, $z$ supposed to be the sides of a valid triangle, it sets 
variable $type$ according to the type of the triangle:
equilateral ($type=2$), isosceles ($type=1$) or scalene ($type=0$).
Figure~\ref{fig:ToyTrityp} also illustrates fourteen test objectives from common test criteria labelled from $l_1$ to $l_{14}$.
$l_1$ and $l_2$ require to cover both possible decisions (or branches) of the conditional at line 6.
For example, covering $l_2$ means to find test data such that during its execution 
the location of $l_2$ is reached and the condition \lstinline'x != y || y != z' is true at 
this location, which ensures to execute the else branch.
Similarly, $l_5$ and $l_{6}$ require to cover both decisions at line 12.  
These four objectives are  specified by the Decision Coverage (DC) criterion for this program.
$l_3$ and $l_4$ (resp., $l_7$ and $l_{8}$) require to cover both truth  values of the first condition 
in the compound condition on line 6 (resp., line 12). 
They are imposed by  Condition Coverage (CC)   
 -- the similar test objectives imposed by CC for the other conditions are not shown to improve readability. 
$l_9$ and $l_{10}$ provide examples of objectives from 
Multiple Condition Coverage (MCC)  for  conditional at line 12. MCC requires to cover all combinations 
of truth values of conditions.
Finally, objectives $l_{11}$ to $l_{14}$ encode some Weak Mutations (WM) of the assignment on line 15
(see \cite[Th.~2]{bardin14} for more detail). 

\smallskip

We can easily notice that $l_9$ and $l_{10}$ put unsatisfiable constraints over $x$, $y$ and $z$. They are thus \emph{infeasible} objectives and trying to cover them would be a waste of time.  
Other objectives are \emph{duplicates}, denoted by $\Leftrightarrow$: 
they are always covered (i.e. reached and satisfied) simultaneously.
We obviously have $l_3\Leftrightarrow l_7$ and $l_4\Leftrightarrow l_8$ since 
the values of $x$ and $y$ do not change in-between.
Although syntactically different, $l_{13}$ and $l_{14}$ are also duplicates, as they are always reached together (we call them \textit{co-reached} objectives) and satisfied if and only if $type\neq 0$.
Finally, we refer to objectives like $l_{11}$ and $l_{12}$ as being \textit{trivial} 
duplicates: they are co-reached, and always satisfied as soon as reached.  
While we do not have $l_1\Leftrightarrow l_5$, covering $l_{1}$ necessarily implies covering $l_{5}$,
that is, $l_{1}$ \emph{subsumes}  $l_{5}$, denoted $l_{1}\Ra l_{5}$. Other examples of subsumed objectives can be found, like $l_{6}\Ra l_{2}$. 
Duplicate and subsumed objectives are redundant objectives that can skew the measurement of test suite strength which should be provided by the test coverage ratio. For example, considering the objectives from the DC criterion, the test suite composed of the single test $(x=1,y=2,z=1)$ covers $l_2$ and $l_5$ but not $l_1$ and $l_6$, which implies a medium coverage ratio of 50\%. The tester may be interested to know the achieved level of coverage without counting duplicate or subsumed objectives. Here, $l_2$ and $l_5$ are actually subsumed by $l_1$ and $l_6$. If the subsumed objectives are removed, the coverage ratio falls down to 0\%. Discarding redundant objectives provides a better measurement of how far testers are from building an efficient test suite, containing only the necessary inputs for covering the non redundant objectives ($l_1$ and $l_6$ in this case). 

\smallskip

The main purpose of this paper is to provide a lightweight yet powerful technique for pruning out 
infeasible, duplicate and subsumed test objectives. To do so, our approach first focuses on infeasible objectives. In Figure~\ref{fig:ToyTrityp}, one can notice, for instance, that the problem of proving $l_9$ to be infeasible can be reduced to the problem of proving that a code assertion \lstinline'!(x!=y && y==z && x==z)' at line 11 will never be violated. Our approach then delegates this proof for each objective to a dedicated verification tool.   
While infeasibility should be checked once per objective, {\it duplication and subsumption require to analyze all the possible pairs}. To avoid quadratic complexity, we focus on detecting duplicate and subsumed pairs 
only among the objectives that belong to the same sequential block of code, 
with no possible interruption of the control flow  (with goto, break, \dots) 
in-between. By construction, the objectives in these groups are always co-reached.
In Figure~\ref{fig:ToyTrityp}, $l_1$--$l_{10}$ and $l_{11}$--$l_{14}$  are two examples of such groups.
Examples of duplicate and subsumed objectives within these groups include
$l_3 \Leftrightarrow l_7$, $l_4 \Leftrightarrow l_8$, $l_{11} \Leftrightarrow l_{12}$, $l_{13} \Leftrightarrow l_{14}$, $l_{1}\Ra l_{5}$
and $l_{6}\Ra l_{2}$.
We call them \emph{block-duplicate} and \emph{block-subsumed} objectives.
On the other hand, $l_1$ and $l_{13}$ are duplicate (at line 14,
\lstinline'type' is nonzero 
if and only if \lstinline'x', \lstinline'y', and \lstinline'z' are equal), but this will not be detected
by our approach since those labels are not in the same block.

%% file: FigToyTrityp.tex
\begin{figure}[tb]
\begin{lstlisting}[basicstyle=\scriptsize\ttfamily,mathescape]
// given three sides x,y,z of a valid triangle, computes 
// its type as: 0 scalene, 1 isosceles, 2 equilateral
int type = 0;  
// l1:  x == y && y == z; (DC) l2:  x != y || y != z; (DC) 
// l3:  x == y; (CC)           l4:  x != y; (CC)
if( x == y && y == z ){
    type = type + 1;
}
// l5: x==y || y==z || x==z; (DC)  l6: x!=y && y!=z && x!=z; (DC)
// l7: x == y; (CC)                l8: x != y; (CC)
// l9: x!=y && y==z && x==z; (MCC) l10: x==y && y!=z && x==z;(MCC) 
if( x == y || y == z || x == z ){
// l11: type + 1 != type + 2; (WM)  l12: type + 1 != type; (WM)
// l13: type + 1 != -type + 1; (WM) l14: type + 1 != 1; (WM)
    type = type + 1;
}
\end{lstlisting}
  \vspace{-4mm}
  \caption{Toy example of a C program with test objectives}
  \vspace{-5mm}
  \label{fig:ToyTrityp}
\end{figure}

%% file: background.tex
\section{Background}
\label{sec:background}


\subsection{Test Objective Specification with Labels}
\label{subsec:backgr-labels}
\vspace{-1mm}



Given a program $\Pg$ over a vector $V$ of $m$ input variables taking values in a domain
$\InputDom \triangleq \InputDom_1 \times \dots \times \InputDom_m$,
a \textit{test datum} $\TD$ for $\Pg$ is a valuation of $V$, i.e.\ $\TD \in \InputDom$. A \textit{test suite} $\TS \subseteq \InputDom$ 
is a finite set of test data. 
A (finite) execution of $\Pg$ over some $\TD$, denoted $\Pg(\TD)$, is a 
(finite) run $\sigma \triangleq \langle(\loc_0,\astate_0),\dots,(\loc_n,\astate_n)\rangle$
where the $\loc_i$ denote successive (control-)locations
of $\Pg$ ($\approx$ statements of the programming language in which $\Pg$
is written), $\loc_0$ refers to the initial program state
and the $\astate_i$ denote the successive internal states of $\Pg$
($\approx$ valuation of all global and local variables and of all memory-allocated structures)
after the execution of each $\loc_i$.

A test datum $\TD$ \textit{reaches} a location $\loc$ at step $k$ 
with internal state $\astate$, denoted $\TD \covers_\Pg^{k} ( \loc, \astate )$, if $\Pg(\TD)$ has the form
\mbox{$\sigma \cdot ( \loc, \astate ) \cdot \rho$}
where $\sigma$ is a partial run of length $k$.  When focusing on reachability, we omit $k$ and write $\TD \covers_\Pg ( \loc, \astate )$. 

Given a test objective {\bf c}, we write  $\dt \covers_P \text{\bf c}$ if test datum $\dt$ covers {\bf c}.
We extend the notation for a test suite $\TS$ and a set of test objectives {\bf C}, writing   $\TS \covers_P \text{\bf C}$
when 
for any $\text{{\bf c}} \in \text{{\bf C}}$, there exists $\dt \in \TS$ such that $ \dt \covers_P \text{\bf c}$. 
%
{A \textit{(source-code based) coverage criterion} $\mathbb{C}$ is defined as a systematic way of deriving a set of test objectives $\text{{\bf C}} \triangleq \mathbb{C}(\Pg)$ for any program under test $\Pg$. 
A test suite $\TS$ satisfies (or achieves) a coverage criterion $\mathbb{C}$ if $\TS$ covers  $\mathbb{C}(\Pg)$.
%



\myparagraph{Labels.}
\textit{Labels} have been introduced in~\cite{bardin14}
as a code annotation language to encode concrete test
objectives. Several common coverage criteria
can be simulated by label coverage, in the sense that for a
given program $P$ and a
criterion $\mathbb{C}$, every concrete test objective from $\mathbf{C} \triangleq \mathbb{C}(P)$
can be encoded using a \textit{corresponding} label. 


Given a program \Pg, a \textit{label} $\Lab \in \Labs$ is a pair $( \loc, \labpred )$ where \loc is a location of P and \labpred is a predicate over the internal state at \loc.
There can be several labels defined at a single location, which can possibly
share the same predicate. More concretely, the notion of labels can be compared to labels
in the C language, decorated with a pure (i.e. side-effect-free)
boolean C expression.

We say that  a test datum $\TD$ \textit{covers a label} $\Lab \mydef (\loc, \labpred )$ in  $\Pg$, denoted $\TD \labcovers_\Pg \Lab$, if there is a state \astate such that \TD 
reaches $(\loc,\astate)$ (i.e.\ $\TD \covers_\Pg ( \loc, \astate )$) and \astate satisfies \labpred. 
An \textit{annotated program} is a pair $\langle \Pg, \LabSet \rangle$ where $\Pg$ is a program and $\LabSet \subseteq \Labs$ is a set of labels for $P$.
Given an annotated program $\langle \Pg, \LabSet \rangle$, we say that a test suite \TS satisfies the \textit{label coverage criterion} (\textbf{LC}) for $\langle \Pg, \LabSet \rangle$, denoted $\TS \labcovers_{\langle \Pg, L \rangle} \mathbf{LC}$, if 
\TS covers every label of \LabSet (i.e. \ $\forall \Lab \in \LabSet : \exists \TD \in \TS: \TD \labcovers_\Pg \Lab$).

\myparagraph{Criterion Encoding.} 
Label coverage \textit{simulates a coverage criterion} $\mathbb{C}$
if any program $P$ can be {\em automatically} annotated with a set of
\textit{corresponding} labels $L$ in such a way that any test suite $TS$ satisfies
$\mathbf{LC}$ for $\langle \Pg, \LabSet \rangle$ if and only if \TS covers
all the concrete test objectives instantiated from $\mathbb{C}$ for
$P$. 
The main benefit of labels is to {\it unify} the treatment of test requirements belonging to 
different classes of coverage criteria  in a transparent way, thanks to the {\it automatic insertion} of labels in the program under test. 
%
Indeed, it is shown in \cite{bardin14}  that label coverage can notably simulate  basic-block coverage (\textbf{BBC}), branch coverage (\textbf{BC}) and decision coverage (\textbf{DC}), 
function coverage (\textbf{FC}), 
condition coverage (\textbf{CC}), decision condition coverage (\textbf{DCC}),  multiple condition coverage (\textbf{MCC})  as well as the  side-effect-free fragment of
 weak mutations (\textbf{WM'}). 
The encoding of \textbf{GACC} comes 
from~\cite{Pandita2010PexMCDC}. 
Some examples are given  in Figure~\ref{fig:ToyTrityp}.

\myparagraph{Co-reached Labels.} 
We say that location $\loc$ is \emph{always preceded by} location $\loc'$
if for any test datum $\TD$, whenever the execution $\Pg(\TD) \triangleq \langle(\loc_0,\astate_0),\dots,(\loc_n,\astate_n)\rangle$
passes through location $\loc$ at step $k$ (i.e. $\loc=\loc_k$) then $\Pg(\TD)$ also passes 
through  $\loc'$ at some earlier step $k'\le k$  (i.e. $\loc'=\loc_{k'}$)
without passing through $\loc$ or $\loc'$ in-between (i.e. at some intermediate step $i$ with $k'< i < k$). 
Similarly, $\loc'$ is said to be \emph{always followed by} location $\loc$
if for any $\TD$, whenever the execution $\Pg(\TD)$
passes through $\loc'$ at step $k'$ then $\Pg(\TD)$ also passes through $\loc$ at some later step $k\ge k'$  
without passing through $\loc$ or $\loc'$ in-between.
Two locations are \emph{co-reached} if one of them is always preceded by the other, 
while the second one is always followed by the first one. 
Note that we exclude the case when one of locations is traversed several times (e.g. due to a loop) 
before being finally followed by the other one.
In a sequential block of code, with no possible interruption of the control
flow  in-between (no goto, break, \ldots), all locations are co-reached.
Finally, two labels are \emph{co-reached} if their locations are co-reached.

\vspace{-1mm}
\subsection{Polluting Labels}
\vspace{-1mm}
\label{subsec:backgr-polluting}

In the remainder of the paper, test objectives will often be expressed in terms of labels.
This work addresses three kinds of polluting labels: infeasible, duplicate and subsumed.
A label $\Lab$ in $\Pg$ is called \emph{infeasible} if there is no test datum $\TD$ 
such that $\TD \labcovers_\Pg \Lab$. In other words, it is impossible to reach its
location and satisfy its predicate.

We say that a label $\Lab$  \emph{subsumes} another label $\Lab'$ (or $\Lab'$ \emph{is subsumed by} $\Lab$) in $\Pg$,
denoted $\Lab\Rightarrow \Lab'$,
if for any test datum $\TD$, if $\TD \labcovers_\Pg \Lab$ then  $\TD \labcovers_\Pg \Lab'$ as well.
Finally, two labels $\Lab$ and $\Lab'$ are called \emph{duplicate}
which in mutation testing  means \emph{infeasible objective}.},
denoted  $\Lab\Leftrightarrow \Lab'$,
if each of them subsumes the other one.
For the specific case where both labels $\Lab$ and $\Lab'$ 
belong to the same group of co-reached labels in a block, 
we call a duplicate (resp., subsumed) label \emph{block-duplicate} (resp., \emph{block-subsumed}).

Notice that if a label $\Lab$ is infeasible, it subsumes by definition any other label  $\Lab'$. 
We call it \emph{degenerate subsumption}. If  $\Lab'$ is feasible, it should be kept and covered.
In this case, the truely polluting objective is $\Lab$ rather than  $\Lab'$. 
That is the reason why it is necessary to eliminate as many infeasible labels as possible before 
pruning out subsumed labels.



\vspace{-1mm}
\subsection{The Frama-C/LTest Platform} 
\label{subsec:backgr-ltest}
\vspace{-1mm}


Frama-C~\cite{FRAMAC} is an open-source industrial-strength framework dedicated to the formal analysis of C programs. 
It has been  successfully used in several safety  and security critical contexts. 
The tool is  written in OCaml, and represents a very significant development (around 150K lines for the kernel and the main plug-ins alone).

Frama-C is based on a small kernel that takes care of providing an abstract representation
of the program under analysis and maintaining the set of properties that are
known about the program state at each possible execution step. These properties
are expressed as ACSL~\cite{ACSL} annotations. 
On top of the kernel, many plug-ins can perform various kinds of analysis, and
can interact with the kernel either by indicating that a property $\phi$ holds,
or by asking whether some other property $\psi$ is true (in the hope that
another plug-in will be able to validate $\phi$ later on).

In the context of this paper, we are mainly interested in the four following
(open-source) plug-ins.
{\it LAnnotate}, {\it LUncov} and {\it LReplay} 
are part of  Frama-C/LTest~\cite{Bardin2014TAP,Marcozzi17b}. 
LAnnotate annotates the program with labels according to the selected criterion. 
LUncov combines weakest-precondition and value analysis to detect infeasible test objectives.
LReplay executes a test suite and computes its coverage ratio.
%
{\it WP} is a plug-in implementing weakest-precondition
calculus~\cite{HoareCommunicationsoftheACM1969,BarnettLeinoPASTE05} in order to
prove that an ACSL assertion holds. 

%% file: approach.tex
\vspace{-1mm}
\section{The LClean approach}
\label{sec:approach}
\vspace{-1mm}

\begin{figure*}[tb]
\includegraphics[trim=0pt 650pt 0pt 0pt, clip,width=\textwidth]{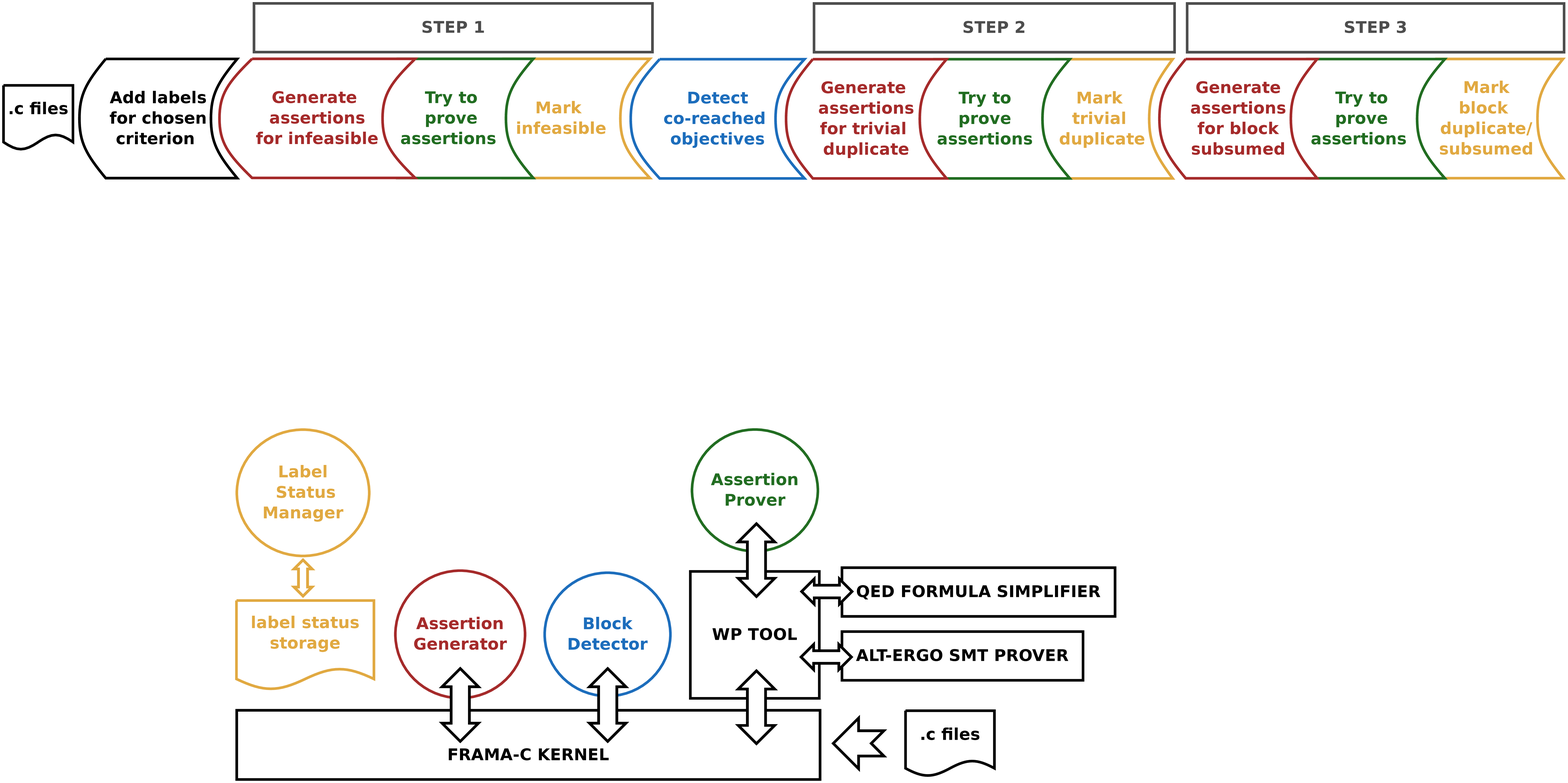}
\vspace{-7mm}
\caption{Process view of the LClean approach with main steps and substeps}
\label{fig:lclean-app}
\vspace{-3mm}
\end{figure*}

The LClean technique contains three main steps (cf. Figure \ref{fig:lclean-app}) preceded by a preprocessing phase. The first step aims at detecting infeasible label-encoded objectives. The second step targets trivial block-duplicate labels, while the third step focuses more generally on block-subsumed and block-duplicate labels. 




Given a program $\Pg$ and a coverage criterion $\mathbb{C}$ that can be simulated
by labels, the preprocessing generates the corresponding
labels $L$. For C programs, this is done by the LAnnotate plug-in of Frama-C.
Our approach operates on the annotated program $\langle \Pg, \LabSet \rangle$
and marks polluting labels so that they can be pruned out.

\vspace{-1mm}
\subsection{Step 1: Infeasible Labels}
\label{subsec:detect-infeasible}
\vspace{-1mm}

LClean systematically explores $\langle \Pg, \LabSet \rangle$ and replaces every label $\Lab \mydef (\loc, \labpred )$
by an assertion \lstinline[mathescape]'assert(!$\labpred$)', whose predicate is the negation of the label condition.
The resulting assertion-laden code is sent to a deductive verification tool designed for proving that the
received program is correct w.r.t. the defined assertions, i.e. that none of them can be violated during
a possible run of the program.
In practice, the verification tool returns the list of the assertions that it was able to prove correct. 
Since each assertion is by construction the negation of a label condition, the corresponding labels 
are formally proven to be infeasible, and are marked as so. 
These marks will be both used as a final result of the approach 
and as internal information transmitted to the next two steps of LClean. 
Regarding Figure~\ref{fig:ToyTrityp}, LClean  indeed detects that $l_9$ and $l_{10}$ are infeasible.

\vspace{-1mm}
\subsection{Detection of Co-reached Labels}
\label{subsec:detect-blocks}
\vspace{-0.5mm}

\input FigBlockDetection.tex
Prior to Steps 2 and 3, LClean performs the detection of blocks of co-reached locations.
We illustrate it using the sample program of Figure \ref{fig:BlockDet}. 
First, a basic syntactic analysis detects six blocks in the program: 
the global block of each of the two functions, the two branches of the outer 
conditional (line 7), and the then branches of the two nested conditionals. 
Second, a call-graph analysis discovers that the first function is only 
called once in the whole program, so that its outer block can be seen as 
executed as a part of the block containing the function call. 
The two blocks can then be merged. 
Finally, a conservative control-flow interruption analysis detects that 
the \lstinline'exit(0);' statement at line 9 may interrupt the control-flow within 
the then branch of the outer conditional. 
The corresponding block is thus split into two blocks, gathering respectively 
the statements before and after the \lstinline'exit(0);' statement. 
The identified blocks allow us to conclude that there are four groups of mutually co-reached labels: 
$\{l_2\}$,  $\{l_3, l_4, l_1\}$, 
$\{l_5,l_6\}$ 
and $\{l_7\}$.


\vspace{-1mm}
\subsection{Step 2: Trivial Block-Duplicate Labels}
\label{subsec:triv-block-dup}
\vspace{-1mm}
As in Step 1, LClean systematically explores $\langle \Pg, \LabSet \rangle$ and replaces labels by assertions. 
Except for the labels marked as infeasible in Step 1, which are simply dropped out, 
each label $\Lab \mydef (\loc, \labpred )$ is replaced
by an assertion \lstinline[mathescape]'assert($\labpred$)'. 
This time, the predicate is directly the label condition. 
The resulting assertion-laden code is sent to the verification tool. The proven 
assertions correspond to labels that will be always satisfied as soon as their location is reached. 
Afterwards, LClean identifies among these always-satisfied-when-reached the groups of co-reached labels 
(cf. Section \ref{subsec:detect-blocks}).
The labels within each of the groups are trivial block-duplicates, and they are marked as being 
clones of a single label chosen among them. 
Again, these marks will be both final results and internal information transmitted to the next step.       
For the example of Figure~\ref{fig:ToyTrityp}, LClean will identify 
that $l_{11}$ and $l_{12}$ are trivial block-duplicate labels.
Similarly, if we assume that all predicates $\labpred_i$ are always satisified
for the code of Figure \ref{fig:BlockDet}, Step 2 detects that $l_3$, $l_4$ and $l_1$ 
are trivial duplicates, and $l_5$ and $l_6$ are as well.
As a subtle  optimization, LClean can detect that label $l_2$ is always executed 
simultaneously with the outer conditional, so that $l_2$ will be covered 
if and only if at least one of the labels $l_3$ and $l_6$ is covered. 
$l_2$ can thus be seen as duplicate with the pair ($l_3$,$l_6$) and is marked as so.


\vspace{-1mm}
\subsection{Step 3: Block-Subsumed Labels}
\label{subsec:block-subsumed}
\vspace{-1mm}

Within each group of co-reached labels, the labels previously detected as infeasible by Step 1 are removed 
and those detected as trivial block-duplicates by Step 2 are merged into a single label. 
Afterwards, every label $\Lab_i = (\loc_i, \labpred_i )$ remaining in the group
is replaced by a new statement \lstinline[mathescape]'int vl$_i$ = $\labpred_i$;', 
which assigns the value of the label condition to a fresh variable \lstinline[mathescape]'vl$_i$'. 
Then, for each pair $(\Lab_i,\Lab_j)_{i\neq j}$ of co-reached labels in the group, the assertion
\lstinline[mathescape]'vl$_j$' \lstinline[mathescape]'$\implies$vl$_j$' is inserted at the end of
the corresponding block of co-reached locations. 
If this assertion is proven by the verification tool, 
then label $\Lab_i$ subsumes label $\Lab_j$. 
Indeed, their locations are co-reached, and the proven assertion shows that 
every input satisfying $\labpred_i$ will also satisfy $\labpred_j$. 
As a consequence, every input that covers $\Lab_i$ also covers $\Lab_j$. 

The graph of subsumption relations detected in a group of co-reached labels is then searched for cycles. 
All labels in a cycle are actually duplicates and can be marked as mergeable into a single label. 
Among the labels that survive such a merging phase, those that are pointed to by at 
least one subsumption relation are marked as subsumed labels. 
For the example of Figure~\ref{fig:ToyTrityp}, LClean will identify,  for instance, 
$l_{1}\Ra l_{5}$,  $l_{6}\Ra l_{2}$, $l_{3}\Leftrightarrow l_{7}$ and  $l_{13}\Leftrightarrow l_{14}$.

\vspace{-1mm}
\subsection{Final Results of LClean}
\label{subsec:lclean-results}
\vspace{-1mm}
Once this third and final step finished, LClean returns a list of polluting labels
composed of the infeasible ones returned by Step 1 and of the duplicate and subsumed
ones returned by Steps 2 and 3.
It should be noted that the approach is modular and each of the three main steps can also be
run independently of the others. However, removing infeasible objectives before Steps 2
and 3 is important, as it reduces the risk of returning degenerate subsumption relations.
Similarly, Step 2 detects duplicate labels that would be identified by Step 3 anyway, 
but Step 2 finds them at much lower cost.
Indeed, the number of proofs required by Step 2 is linear in the number of labels 
as it does not have to consider pairs of labels.

The LClean approach might be extended to detect duplicate or subsumed labels that are not
in the same basic block, by generating more complex assertions that would be flow-sensitive.
However, limiting the analysis to block-duplicate and block-subsumed labels
turns out to be a sweet spot between detection power and computation time. 
Indeed, Figure \ref{fig-n} details the total number of  pairs of labels for four common criteria
in the 14 C programs used in the evaluation in Section \ref{sec:evaluation}  (cf. Figure \ref{fig-all}). 
Figure \ref{fig-n} also presents the total number of pairs of labels taken inside the same block,
inside the same function or over the whole program.
We can see that focusing the analysis on block pairs enables reducing the number of necessary
proofs by one to four orders of magnitude.
At the same time, it seems reasonable to think that a significant part of the duplicate or
subsumed labels reside within the same basic block, as those labels are always executed
together and typically describe test objectives related to closely interconnected
syntactic elements of the program. 
{\footnotesize
\begin{figure}[tb]
\centering
\begin{tabular}{|c|c||c||c|c|}
\hline
{\textbf{Criterion}} & {\textbf{Labels}}  & \textbf{{Block Pairs}} & \textbf{{Function Pairs}} & {\textbf{Program Pairs}}  \\
\hline
CC & 27,638 & 94,042 & 3,013,940 & 428,075,244 \\
MCC & 30,162 & 314,274 & 3,961,004 & 503,856,852 \\
GACC & 27,638 & 94,042 & 3,013,940 & 428,075,244 \\
WM & 136,927 & 2,910 908 & 80,162,503 & 8,995,885,473 \\
\hline
\hline
\multirow{2}{*}{\textbf{TOTAL}} & \textbf{222,365} & \textbf{3,413,266} & \textbf{90,151,387} & \textbf{10,355,892,813} \\
 & $\mathbf{\color{red}(\times 1/15)}$ & $\mathbf{\color{red}(\times 1)}$ &  $\mathbf{\color{red}(\times 26)}$ & $\mathbf{\color{red}(\times 3034)}$  \\
\hline
\end{tabular} 
\vspace{-4mm}
\caption{Number of pairs of labels in 14 C programs}
\label{fig-n}
\vspace{-4mm}
\end{figure}}

%% file: FigBlockDetection.tex
\begin{wrapfigure}[19]{l}{27mm}
\vspace{-4mm}
%
%
\begin{lstlisting}[basicstyle=\scriptsize\ttfamily,mathescape]
void calledOnce () { 
  // l1: $\labpred_1$
  code1; 
}
int main ( int i ) {
// l2: $\labpred_2$
if (i>0) {
  // l3: $\labpred_3$
  if (i==5) i++; 
  // l4: $\labpred_4$
  calledOnce ();
  if (i==7) exit(0);
  // l5: $\labpred_5$
  i++;
  // l6: $\labpred_6$
} else { 
  // l7: $\labpred_7$
  code2; 
}
return i;
} 
\end{lstlisting} 
\vspace{-5mm}
\caption{Co-reached locations}
\vspace{-4mm}
\label{fig:BlockDet}
\end{wrapfigure}

%% file: implementation.tex
\vspace{-1mm}
\section{Implementation}
\label{sec:implementation}
\vspace{-1mm}

\begin{figure}[b]
\vspace{-3mm}
\includegraphics[trim=250pt 0pt 480pt 470pt, clip,width=\columnwidth]{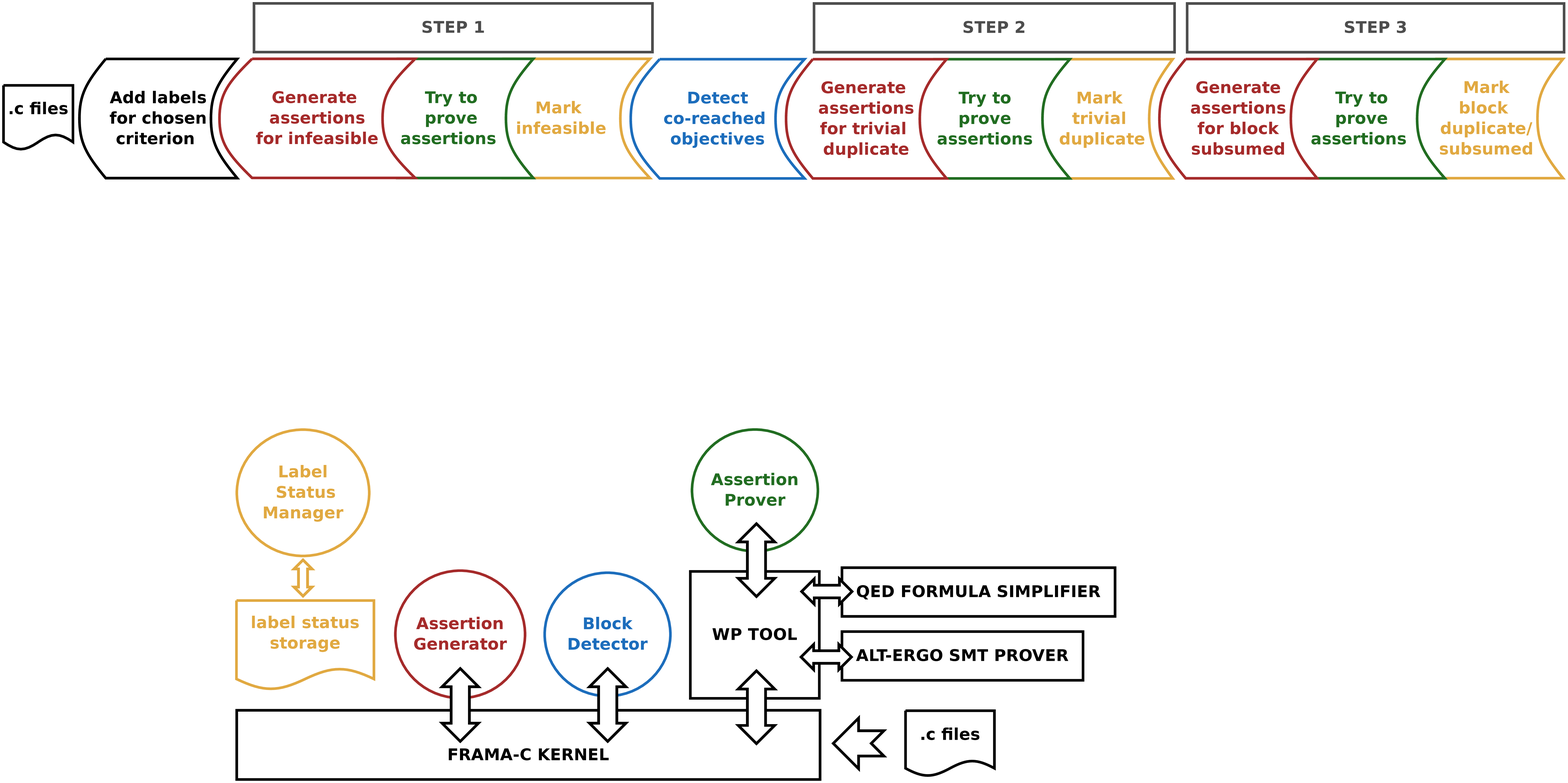}
\vspace{-6mm}
\caption{Frama-C plug-in implementing one LClean step}
\label{fig:lclean-tool}
\end{figure}

The LClean approach is implemented in \emph{three independent open-source Frama-C plug-ins}\footnote{\label{foot:website}
Available from \url{https://sites.google.com/view/lclean}.} written in OCaml ($\approx$5,000 locs).
These plug-ins share a common architecture depicted in Figure \ref{fig:lclean-tool}. 
It relies on the Frama-C kernel (in black) 
and features four modules (in color) performing the different substeps 
of an LClean step. 
It receives as input an annotated program $\langle \Pg, \LabSet \rangle$, in which
labels have already been generated with plug-in LAnnotate \cite{Bardin2014TAP} 
in order to simulate the coverage criterion of interest. As a starting point,
the program 
is parsed by the Frama-C kernel, which makes its abstract syntax tree (AST) 
available for all the components of the architecture. We now
present the four modules performing the analysis.

\myparagraph{Assertion Generator.}
The Assertion Generator replaces the labels in the code 
by assertions according to the corresponding step (cf. Section \ref{sec:approach}). 
Frama-C primitives are used to explore the AST, locate the nodes corresponding to labels and 
replace them by the required assertions, written in ACSL.


\myparagraph{Robust Multicore Assertion Prover.} \label{sec:optims}
The Assertion Prover deals with proving the assertions introduced in the 
AST by the Assertion Generator and relies on the WP plug-in. 
It is not a simple wrapper for WP:
the Assertion Prover introduces {\it crucial optimizations 
ensuring its scalability and robustness}: 
\begin{itemize} 
\item First, it embeds a version of WP that we carefully optimized 
for our specific needs, making it capable to prove several different 
assertions independently in a single run of the tool. 
This version factors out a common part of the analysis 
(related to the program semantics) that would have 
to be repeated uselessly if WP was called once per assertion. 
\item Second, 
its multi-core implementation ensures a significant speedup.
The assertions to be proved are shared 
among several parallel WP instances running on different cores. 
%
%
%
\item Third, the Assertion Prover 
also guarantees robustness and adaptability of the process. 
Indeed, the WP tool can consume a high amount of memory and 
computation time when analyzing a large and complex C function. 
The Assertion Prover can smoothly interrupt a WP session
when a threshold  w.r.t. the used memory or elapsed time has been reached.  
\end{itemize} 

All these improvements to Frama-C/WP have been proven  crucial for large-scale 
experiments (cf.~Section \ref{sec:evaluation}).   

\myparagraph{Label Status Manager.}
The Label Status Manager maintains and gives access to a set of files 
storing a status for each label. 
Each label is identified by a unique integer ID used both in the AST and in the status files. 
The status of a label can be a) infeasible, b) 
duplicate to another ID (or a pair of IDs), c) subsumed by other IDs, or d) unknown. 
The status files are updated by the plug-ins when they detect 
that some labels can be marked as polluting. 
The plug-ins for Steps 2 and 3 also check the files in order 
to drop out 
the labels marked as polluting during the previous steps.

\myparagraph{Block Detector.}
The detector of blocks of co-reached labels is only used before Steps 2 and 3. 
It relies on the Frama-C primitives to explore the AST and 
perform the analyses detailed in Section \ref{subsec:detect-blocks}.
For each block found, it returns the label IDs of co-reached labels belonging to the block.

%% file: evaluation.tex
\vspace{-2mm}
\section{Experimental Evaluation}
\label{sec:evaluation}
\vspace{-1mm}

To evaluate experimentally  LClean, we consider the following three research questions:

\textbf{Research Question 1 (RQ1)}: Is the approach effective and useful? Especially, (a) Does it identify  a significant number of objectives from common criteria, 
all being real polluting objectives? b) Can it scale to real-world applications, involving many lines of code and complex language constructs? 

\textbf{Research Question 2 (RQ2)}: Do the optimizations (Section \ref{sec:optims})   improve the time performance in a significant way, impacting LClean acceptability in practice?

\textbf{Research Question 3 (RQ3)}: How does our approach compare with the closest approaches like LUncov, mutant classification and TCE, especially in terms of pruning power and time performance? 

The experimental artefacts used to answer these questions and the fully detailed results that we obtained are available on the companion website$^{\ref{foot:website}}$ of the paper. The tool and artefacts have also been installed in a Linux virtual machine provided on the website and enabling an easy reproduction of the experiments  described in the next subsections. All these experiments were performed on a Debian Linux 8 workstation equipped with two Intel Xeon E5-2660v3 processors, for a total of 20 cores running at 2.6Ghz and taking advantage of 25MB cache per processor and 264GB RAM.

\vspace{-1mm}
\subsection{RQ1: Effectiveness and Scalability}
\label{subsec:rq1}
\vspace{-1mm}

We consider fourteen C programs of various types and sizes (min: 153 locs, mean: 16,166 locs, max: 196,888 locs) extracted from five projects: the seven Siemens programs from \cite{siemensTS}, four libraries taken from the cryptographic OpenSSL toolkit \cite{openssl}, the full GNU Zip compression program \cite{gzip}, the complete Sjeng chess playing IA application \cite{sjeng} and the entire SQLite relational database management system \cite{sqlite}. Every program is annotated successively with the labels encoding the test objectives of four common coverage criteria: Condition Coverage (CC), Multiple-Condition Coverage (MCC), General Active Clause Coverage (GACC) and Weak Mutations (WM, with sufficient mutation operators \cite{offutt96}). The LClean tool is then run to detect polluting objectives for each (program, criterion) pair. 

\medskip

For each step of the LClean process, the number of marked objectives and the computation time are reported in Figure \ref{fig-all}. 11\% of the 222,365 labels were marked as polluting in total (min: 4\% for CC/MCC with SQLite, max: 27\% for WM in Siemens/printokens.c). The global ratio of marked polluting objectives is 5\% for CC, 5\% for MCC, 6\% for GACC and 15\% for WM. In total, 13\% of the detected polluting objectives were infeasible, 46\% were duplicate (about one half was marked during Step 2 and the other during Step 3) and 41\% were subsumed. The computation time ranges from 10s for MCC in Siemens/schedule.c (410 locs and 58 objectives) to $\sim$69h for WM in SQLite (197K locs and 90K objectives). Globally, computation time is split into 10\% for Step 1, 8\% for Step 2 and 82\% for Step 3. 
While the computation time is acceptable for a very large majority of the experiments, Step 3 becomes particularly costly when applied on the largest programs, with the most meticulous criteria. 
This is of course due to the fact that this step is quadratic in the number of
labels. While we limit our analysis to block pairs, the number of resulting proof attempts still gets large for bigger applications, reaching 1.8M proofs for SQLite and WM (which remains tractable). 
Yet,  limiting LClean to Steps 1 \&  2 still marked many labels and is much more tractable: on SQLite, it detects 4566 polluting objectives in only 9h 
(13692 objectives in 69h for full LClean).  



{\it Conclusion. These results indicate that LClean is an effective and useful approach able to detect that a significant proportion of the test objectives from various common criteria are (real) polluting ones, even for  large and complex real-word applications.  
In practice, for very large programs and demanding criteria,  LClean can be limited to Steps 1 \& 2,  keeping a significant detection power at a much lower expense.  
}


{\footnotesize
\begin{figure*}[tb]
\centering
\begin{tabular}{|c|c|c|c|c|c|c|c|c|c|c|c|c|}
\hline 
\multirow{3}{*}{\textbf{Benchmark}} & \multirow{3}{*}{\textbf{Labels}} & \multicolumn{2}{c|}{\textbf{STEP 1}}  & \multicolumn{2}{c|}{\textbf{STEP 2}} & \multicolumn{3}{c|}{\textbf{STEP 3}}  & \multicolumn{3}{c|}{\textbf{TOTAL}} & \multirow{3}{*}{\textbf{Criterion}} \\
\cline{3-12}
 & & \textit{\scriptsize marked as} & \multirow{2}{*}{\textit{\scriptsize time}} & \textit{\scriptsize marked as} & \multirow{2}{*}{\textit{\scriptsize time}} & \textit{\scriptsize marked as}  & \textit{\scriptsize marked as} & \multirow{2}{*}{\textit{\scriptsize time}} & \multicolumn{2}{c|}{\textit{\scriptsize marked as polluting}} & \multirow{2}{*}{\textit{\scriptsize time}} &  \\ 
\cline{10-11}
  & & \textit{\scriptsize infeasible} &  & \textit{\scriptsize duplicate} & & \textit{\scriptsize duplicate}  & \textit{\scriptsize subsumed} & & \textit{\scriptsize ratio} & \textit{\scriptsize \%} & &  \\ 
\hline 
{\textbf{siemens}}  & 654 & 0 & 35s & 0 & 38s & 2 & 41    
& 83s & 43/654 & {7\%} & {156s} & {CC} \\
\scriptsize {(agg. 7 programs)} & 666 & 20 & 36s & 0 & 40s & 0 & 16    
& 78s & 36/666  &  {5\%} & {154s}  & {MCC} \\ 
\scriptsize {\textit{3210 locs}} & 654 & 1 & 37s & 0 & 39s & 18 &   
 17 & 77s & 36/654 &  {6\%} & {153s}  & {GACC} \\ 
 & 3543 & 37 & 114s & 123 & 126s & 134 & 336 
& 723s & 630/3543  &  {18\%} & {963s}   & {WM} \\  
\hline 
{\textbf{openssl}} 
 & 1022 & 28 & 67s & 3 & 67s & 4 & 57 & 391s & 92/1022 & 9\% & 525s & CC \\
\scriptsize {(agg. 4 programs)}   & 1166 & 134 & 77s & 0 & 83s & 2 & 24 & 294s & 160/1166 & 14\% & 454s & MCC \\ 
\scriptsize {\textit{4596 locs}}  & 1022  & 29 & 70s & 0 & 81s & 30 & 24 & 324s & 83/1022 & 8\% & 475s  & GACC \\ 
 & 4978 & 252 & 356s & 270 & 372s & 200 & 326  
& 4214s &  1048/4978 & 21\% & 5122s  & WM \\ 
\hline 
{\textbf{gzip}}  & 1670 & 23 & 149s & 5 & 152s & 19 & 54   
& 578s & 101/1670 & 6\% & 879s & CC \\
\scriptsize {\textit{7569 locs}} & 1726 & 44 & 170s & 5 & 171s & 11 & 34    
& 628s & 94/1726 & 5\% & 969s & MCC \\ 
 & 1670 & 31 & 154s & 5 & 156s  & 43 & 34   
& 555s & 113/1670 & 7\% & 865s & GACC \\ 
 & 12270 & 267 & 1038s & 942 & 1210s & 542 & 895 
& 10029s &  2646/12270 & 22\% & 12277s   & WM \\ 
\hline 
{\textbf{sjeng}}  & 4090 & 34 & 351s & 15 & 354s   & 82   
& 215 & 798s & 346/4090 & 8\% & 1503s & CC \\
\scriptsize {\textit{14070 locs}} & 4746 & 358 & 417s & 9 & 436s & 34 & 26    
& 1912s & 427/4746 & 9\% & 2765s & MCC \\ 
 & 4090 & 35 & 349s & 15 & 353s & 82 & 210  
& 751s & 342/4090 & 8\% & 1453s & GACC \\ 
 & 25722 & 353 & 5950s & 483 & 4791s & 640 & 706 
& 19586s &  2182/25722 & 8\% & 31478s & WM \\ 
\hline 
{\textbf{sqlite}}  & 20202 & 120 & 1907s & 3 & 1416s & 130 & 456  
& 4646s & 709/20202 & 4\% & 7969 &  CC  \\
\scriptsize {\textit{196888 locs}} & 21852 & 394 & 2295s & 0 & 1902s & 178     
& 255 & 11958s & 827/21852 & 4\% & 16155 &  MCC  \\ 
 & 20202 & 129 & 2065s & 0 & 1613s & 803 & 223  
& 4773s &  1155/20202 & 6\% & 8451 &  GACC \\ 
 & 90240 & 878 & 18104s & 3688 & 13571s &  
 2962 
& 6164 
& { 216140s} 
& 13692/90240 & 15\% & 247815s & WM \\ 
\hline
\hline  
{\textbf{TOTAL}} & 27638 & 205 & 2509s & 26 & 2027s & 237 & 823 & 6496s & \textbf{1291/27638} & \textbf{5\%} & \textbf{3h3m52} &  \textbf{CC} \\
\scriptsize {\textit{226333 locs}}    
& 30156 & 950 & 2995s & 14 & 2632s & 225 & 355 & 14870s & \textbf{1544/30156} & \textbf{5\%} & \textbf{5h41m37} &  \textbf{MCC} \\ 
& 27638 & 225 & 2675s & 20 & 2242s & 976 & 508 & 6480s & \textbf{1729/27638} & \textbf{6\%} & \textbf{3h9m57} &  \textbf{GACC}  \\
& 136753 & 1787 & 25562s & 5506 & 20070s & 4478 & 8427
& 250692s & \textbf{20198/136753} & \textbf{15\%} & \textbf{82h18m44} &  \textbf{WM}  \\
\cline{2-13} 
 & \textbf{222185} & \textbf{3167} & \textbf{9h22m21} & \textbf{5566} & \textbf{7h29m31} & \textbf{5916} & \textbf{10113} & \textbf{77h22m18} & \textbf{ 24762/222185}  &  \textbf{11\%} & \textbf{94h14m10}   & \textbf{TOTAL} \\ 
\hline 
\end{tabular} 
\vspace{-3mm}
\caption{Pruning power and computation time of LClean over 14 various "real-world" C programs}
\label{fig-all}
\vspace{-3mm}
\end{figure*}}

\vspace{-1mm}
\subsection{RQ2: Impact of Optimizations}
\label{subsec:rq2}
\vspace{-1mm}

We repeat the experiments performed in \textbf{RQ1} for the WM criterion over the seven Siemens programs, but we deactivate the optimizations that we implemented in 
the Assertion Prover of our tool, namely  tailored WP tool and  multi-core implementation (Section \ref{sec:optims}). Figure \ref{fig-impact-optims} details the obtained computation times ({\it in logarithmic scale}) for the three steps of the LClean process, considering three levels of optimizations. At level 0 (blue), the Assertion Prover uses a single instance of the classical Frama-C/WP running on a single core. At level 1 (red), the Assertion Prover uses 20 instances of the classical version WP running on 20 cores. Level 2 (beige) corresponds to the actual version of the tool used in \textbf{RQ1}, when all the optimizations are activated: the Assertion Prover uses 20 instances of our tailored version WP running on 20 cores.

{\footnotesize
\begin{figure}[tb]
\begin{center}
\pgfplotsset{every axis legend/.append style={
at={(0.02,0.65)},
anchor=south west}}
\noindent\begin{tikzpicture}[baseline]
\begin{axis}[width=8cm,height=5.4cm,xbar=2pt,ybar=2pt,symbolic x coords={x,STEP 1,y,STEP 2,z,STEP 3,a,TOTAL,b}, xtick=data,
xmin=x,
xmax=b,
ylabel={computation time (s) { in log. scale}},
y label style={at={(axis description  cs:0.1,0.5)},anchor=south},
ymax=1000000,
ymode=log,
ymin=1,
legend entries={1 core + classic WP,20 cores + classic WP,20 cores + tailored WP},
]
\addplot coordinates {(STEP 1,5048) (STEP 2,5096) (STEP 3,33546.0) (TOTAL,43690.0)};  
\addplot coordinates {(STEP 1,302) (STEP 2,318) (STEP 3,1731.0) (TOTAL,2351.0)}; 
\addplot coordinates {(STEP 1,114) (STEP 2,126) (STEP 3,723) (TOTAL,963)};  
\end{axis} \end{tikzpicture}
\end{center}
\vspace{-4mm}
\caption{Tool optimization impact (Siemens, WM)}
\label{fig-impact-optims}
\vspace{-2mm}
\end{figure}
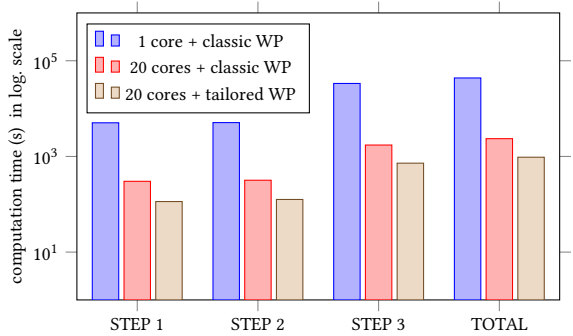}

We observe that the total computation time is reduced by a {\it factor of  2.4} when switching from level 1 to level 2, and that it is   reduced by a {\it factor of 45} when switching from level 0 to level 2. These factors are very similar for all the steps of the LClean process.

{\it Conclusion. These results show that our optimizations have a very significant impact over the time performance of our tool, making the experiments on large programs intractable without them. The measured speedup of 45x has a sensible influence over the perceived speed of the tool, improving its acceptability in practice.} 


\vspace{-1mm}
\subsection{RQ3: LClean vs. Closest Related Work}
\label{subsec:rq3}
\vspace{-1mm}

\subsubsection{LUncov} We apply both  LUncov \cite{bardin15} and LClean  on the same benchmarks \cite{bardin15}.  The measured computation time and detection power for LUncov and LClean are compared in Figure \ref{fig-luncov}. As LUncov is limited to infeasibility, we also provide results for Step 1 of LClean. 
It appears that LClean  detects 4.2$\times$ more polluting labels than LUncov in 1.8$\times$ less time. When LClean is limited to Step 1,  it detects 1.6$\times$ less polluting labels 
than LUncov, but in 10$\times$ less time.

{\it Conclusion. LClean  provides a much more extensive detection of polluting test objectives than LUncov (especially because it goes beyond infeasibility) at a cheaper cost, thanks to its modular approach and optimized implementation.} 


{\footnotesize
\begin{figure}[tb]
\centering
\begin{tabular}{|c|c|c||c|c|c|c|}
\hline 
 \multirow{2}{*}{\textbf{Criterion}} & \multicolumn{2}{|c||}{\textbf{LUncov}} & \multicolumn{2}{|c|}{\textbf{LClean} (step 1)}  & \multicolumn{2}{c|}{\textbf{LClean} (all steps)}  \\
 \cline{2-7}
   &   \textit{marked} & \textit{time} & \textit{marked} & \textit{time} & \textit{marked} & \textit{time}  \\ 
\hline
CC & 4/162 & 97s & 4/162 & 12s & 51/162 & 46s \\
MCC & 30/203 & 125s & 30/203 & 15s & 51/203 & 53s \\
WM & 84/905 & 801s & 41/905 & 75s & 385/905 & 463s \\
\hline
\hline
\multirow{2}{*}{\textbf{TOTAL}} & \textbf{9\%} 
& \textbf{17m3s} 
&  \textbf{6\%}  
& \textbf{1m42s} 
& \textbf{38\%} 
&\textbf{9m22s}   \\
& $\mathbf{\color{red}(\times 1)}$
& $\mathbf{\color{blue}(\times 1)}$
& $\mathbf{\color{red}(\div 1.6)}$ 
& $\mathbf{\color{blue}(\div 10)}$ 
& $\mathbf{\color{red}(\times 4.2)}$  
& $\mathbf{\color{blue}(\div 1.8)}$ 
  \\
\hline
\end{tabular} 
\vspace{-2mm}
\caption{LUncov \cite{bardin15} vs LClean (benchmarks from \cite{bardin15})}
\label{fig-luncov}
\vspace{-4mm}
\end{figure}}

\subsubsection{Mutant Classification} The core principle of mutant classification \cite{SchulerDZ09,SchulerZ13} is to rely on dynamic coverage data to identify  (in an approximated way) polluting mutants. 
As a comparison between LClean and such a   dynamic pruning principle, Figure \ref{fig-dynamic} reveals that the time necessary to run a high-coverage test suite (Siemens test suite), 
save coverage data 
and find  likely-polluting objectives can be one order of magnitude higher than running LClean over the same test objectives. 
In the same time, it appeared that many of  the objectives detected in this way were false positives, leading to a 89\% rate of labels to be considered as likely polluting (mainly because of duplication and subsumption). Actually, while the Siemens test suite  achieves high coverage of standard metrics, it is not built to reveal different coverage behaviours between feasible test objectives. 
Crafting new test cases to do so would reduce the number of false positives but even more penalize the computation time. 


{\it Conclusion. By relying on lightweight static analyses, LClean  provides a sound and quick detection of a significant number of both infeasible and redundant test objectives, while dynamic detection is expensive and unsound, yielding many false positives even based on  high-quality test suites.}


{\footnotesize
\begin{figure*}[tb]
\centering
\begin{tabular}{|c|c|c|c|c|c||c|c|c|c|c|}
\hline 
\multirow{3}{*}{\textbf{Criterion}} & \multicolumn{5}{|c||}{\textbf{Dynamic Detection}} & \multicolumn{5}{|c|}{\textbf{LClean}} \\
\cline{2-11}
& \textit{possibly} & \textit{possibly} &  \textit{possibly} & \textbf{\textit{total ratio for}} & \multirow{2}{*}{\textbf{\textit{time}}} & \textit{marked as} & \textit{marked as} &  \textit{marked as} & \textbf{\textit{total ratio for}} & \multirow{2}{*}{\textbf{\textit{time}}}  \\
& \textit{infeasible} & \textit{duplicate} & \textit{subsumed} & \textbf{\textit{possibly polluting}} & &\textit{infeasible} & \textit{duplicate} & \textit{subsumed} & \textbf{\textit{marked as polluting}}& \\
\hline 
CC & 37/654 & 243/654 & 230/654 & 80\% (510/654) & 3132s & 0/654 & 2/654 & 41/654 & 7\% (43/654) & 156s \\
MCC & 76/666 & 221/666	 & 215/666 & 77\% (512/666) & 3142s & 20/666 & 0/666 & 16/666 & 5\% (36/666) & 154s \\
GACC & 46/654 & 249/654 & 212/654 & 78\% (507/654) & 3134s & 1/654 & 18/654 & 17/654 & 6\% (36/654) & 153s \\
WM & 386/3543 & 2327/3543 & 641/3543 & 95\% (3354/3543) & 8399s & 37/3543 & 257/3543 & 336/3543 & 18\% (630/3543) & 963s \\
\hline
\hline
\multirow{2}{*}{\textbf{TOTAL}} & \multirow{2}{*}{545/5517} & \multirow{2}{*}{3040/5517} & \multirow{2}{*}{1298/5517} & \multirow{2}{*}{\textbf{89\%} (4883/5517)} & \textbf{4h56m47} & \multirow{2}{*}{58/5517} & \multirow{2}{*}{277/5517} & \multirow{2}{*}{410/5517} & \multirow{2}{*}{\textbf{14\%} (745/5517)} & \textbf{23m46s}  \\
 & &  &  &  & $\mathbf{\color{red}(\times 12)}$ &  &  &  &  & $\mathbf{\color{red}(\times 1)}$  \\
\hline 
\end{tabular} 
\vspace{-4mm}
\caption{Dynamic detection of (likely) polluting objectives  vs. LClean (Siemens)}
\label{fig-dynamic}
\vspace{-4mm}
\end{figure*}}

\subsubsection{Trivial Compiler Equivalence (TCE)} A direct comparison with TCE \cite{Papadakis15} is not possible, as TCE aims at identifying strong mutant equivalences, which are fundamentally different from the structural ones we handle.  
This is because the killing of the mutants require the propagation of corrupted program states to the program output, which is  complex to formalize \cite{DeMilloO91}. Thus, the only way to compare the two approaches is to assume that weakly polluting mutants are also strongly polluting ones. 
This assumption is true for the case of equivalent mutants but not entirely true for the case of the duplicate mutants. 
Weakly duplicate mutants might not be strongly duplicate ones as some might turn out to be equivalent due to failed error propagation. However, this is usually quite rare, as most weakly killed mutants propagate to the program output~\cite{OL-94}.  
Nevertheless, we report these results for demonstrating the capabilities of the approaches and not for suggesting a solution for the strong mutation.

To perform the comparison, we generated some strong mutants as well as our corresponding weak ones for the replace program. We selected only the replace program as our purpose here is to demonstrate the relative differences of the approaches: replace is one of the largest program from the Siemens suite, for which TCE performs best with respect to equivalent mutant detection \cite{7882714}. 
Our results show that among the 1,579 mutants involved, our approach detected 103 (7\%) as infeasible,  while TCE detected 96 (6\%). Among these, 91 are shared, which means that 12 of the infeasible mutants were only found by our approach and 5 only by TCE. 
Regarding  duplicate mutants, our approach detected 555 (35\%) as duplicate, and TCE detected 352 (22\%). In this case, 214 were shared, which means that both techniques together identify 693 (44\%) duplicate mutants. 

{\it Conclusion. 
Overall, the results show that our approach outperforms TCE in terms of detection power and  form a relatively good complement of it. Moreover, LClean is able to detect subsumption. 
Yet, TCE is much more efficient, relying on  compiler optimizations. 
}

%% file: threats.tex
\vspace{-2mm}
\section{Discussion}
\label{sec:threats}
\vspace{-1mm}

\subsection{Threats to validity}
\label{subsec:threats-threats}
\vspace{-1mm}

Common to all studies relying on empirical data, this one may be of limited generalizability. To diminish this threat we used in addition to benchmark programs, 5 large real-world ones composed of more than 200 kloc (in total). We also show that our approach can effectively handle real-world programs, like sqlite, and that it is capable of dealing with many types of polluting objectives, that no other approach can handle. 

Our results might also have been affected by the choice of the chosen test criteria  and in particular the specific mutation operators we employ. To reduce this threat, we used popular test criteria (CC, MCC, GACC and WM)  included in software testing standards \cite{reid:standard, radio:standard}, and employed commonly used mutation operators   included in  recent work \cite{ChekamPTH17, AmmannDO14}.

The validity of our experimental results have been crosschecked in several ways. First, we compare our results on the Siemens benchmark  with those of other tools, namely  LUncov and TCE. 
We know by design that infeasible objectives detected by LClean should be detected by LUncov as well, and we check manually the status of each  duplicate objective reported by LClean and not by TCE. No issue was reported.  
%
%
%
%
Second, we used the tests of   the Siemens Suite as a sanity check for redundancy,  
by checking  that every objective reported as infeasible (resp.~duplicate, subsumed) by LClean is indeed  seen as infeasible (resp.~duplicate, subsumed) in the test suite.  
These test suites are extremely thorough \cite{HutchinsFGO94,PapadakisDT14} 
  and are thus likely to detect errors in LClean.    
%
%
%
%
%
%
Third, for larger programs, we pick a random selection of a hundred test objectives reported as infeasible, duplicate or subsumed by LClean and manually check them -- this was often straightforward 
due to the local reasoning of LClean.  All these sanity checks succeeds.  


Another class of threats may arise because of the prototypes we used, as it is likely that Frama-C or our implementation are defective. However, Frama-C is a mature tool with industrial applications in 
highly demanding fields (e.g., aeronautics)  and thus, it is unlikely to  cause important problems. Moreover, our sanity checks   would have likely spotted such issues.


Finally, other threats may be due to the polluting objectives we target. However, infeasible objectives are a well-known issue, usually acknowledged in the literature as one of the most time consuming tasks of the software testing process \cite{SchulerZ13, KurtzAOK16, Papadakis15, ammann08},  and redundant objectives have been stated as a major problem in both   past  and  recent literature~\cite{KurtzAOK16, Papadakis16, KurtzAODKG16}.

\vspace{-1mm}
\subsection{Limitations}
\label{subsec:threats-limitations}
\vspace{-1mm}

Labels cannot address all white-box criteria. For example, dataflow criteria
or full MCDC require additional expressive power~\cite{Marcozzi17a}. Currently, parts of the infeasibility results from LClean could be lifted to these classes of objectives.
On the other hand, it is unclear how it could be done for duplication or subsumption.   
 Extending the present work to these criteria is an interesting direction.

From a more technical point of view, the detection of  subsumption is limited  more or less to basic blocks. While it already enables catching many cases, it
might be possible
 to slightly extend the search while retaining scalability. 
In the same vein, the proofs are performed in LClean on a {\it per} function basis. This is a problem 
as it is often the case that a given function is always 
called within the same context, reducing its possible behaviors. Allowing a limited degree of contextual analysis (e.g., inlining function callers and/or callees) should allow 
to detect more polluting objectives while retaining scalability.  

Finally, as we are facing an undecidable problem, our approach is sound, but not complete: SMT solvers might answer {\it unknown}. In that case, we may miss
polluting objectives.

%% file: related.tex
\vspace{-2mm}
\section{Related Work}
\label{sec:related}
\vspace{-1mm}

\subsection{Infeasible Structural Objectives}
\label{subsec:related-infeas}
\vspace{-1mm}


Early research studies set the basis for identifying infeasible test objectives using constraint-based techniques \cite{OffuttP97, GoldbergWZ94}. Offutt and Pan~\cite{OffuttP97} suggested transforming the programs under test as a set of constraints that encode the test objectives. Then, by solving these constraints, it is possible to identify infeasible objectives (constraints with no solution) and test inputs. Other attempts use model checking \cite{BeyerHJM07, BeyerCM04} to prove that specific structural test objectives (given as properties) are infeasible. Unfortunately, constraint-based techniques, as they require a complete program analysis, have the usual problems of the large (possibly infinite) numbers  of involved paths, imprecise handling of program aliases~\cite{Kosmatov08} and the handling of non-linear constraints~\cite{AnandBCCCGHHMOE13}. Model checking faces precision problems because of the system modelling and scalability issues due to the large state space involved. On the other hand, we rely on a modular, hence not too expensive, form of weakest precondition calculus to ensure  scalability. 

Perhaps the closest work to ours are the ones by Beckman \etal~\cite{BeckmanNRSTT10}, Baluda \etal~\cite{BaludaBDP10, BaludaBDP11, BaludaDP16} and Bardin \etal \cite{bardin15} that rely on weakest precondition. Beckman \etal proves infeasible program statements, Baluda \etal infeasible program branches and Bardin \etal infeasible structural test objectives. Apart from the side differences (Beckman \etal targets formal verification, Baluda \etal applies model refinement in combination to weakest precondition and Bardin \etal combines weakest precondition with abstract interpretation) with these works, our main objective here is to identify all types of polluting test objectives (not only infeasible ones) for real-world programs in a generic way, i.e. for most of the test criteria, including advanced ones such as multiple condition coverage and weak mutation. Another concern regards the scalability of the previous methods, which remains unknown under the combinatorial explosion of test objectives that mutation criteria introduce. 

Other techniques attempt to combine infeasible test objectives detection techniques as a means to speed-up test generation and refine the coverage metric. Su \etal \cite{SuFPHS15} combines symbolic execution with model checking to generate data flow test inputs. Baluda \etal~\cite{BaludaDP16} combines backward (using weakest precondition) and forward symbolic analysis to support branch testing and Bardin \etal \cite{bardin14,bardin15} combines weakest precondition with dynamic symbolic execution to support the coverage of structural test objectives. Although integrating such approaches with ours may result in additional benefits, our main objective here is to demonstrate that lightweight symbolic analysis techniques, such as weakest precondition, can be used to effectively tackle the general problem of polluting objectives for almost all structural test criteria in real-world settings.  

Another line of research attempts diminishing the undesirable effects of infeasible paths in order to speed-up test generation.  Woodward \etal~\cite{WoodwardHH80} suggested using some static rules called allegations to identify infeasible paths. Papadakis and Malevris~\cite{PapadakisM12} and Lapierre \etal~\cite{LapierreMSAFT99} used a heuristic based on the k-shortest paths in order to select likely feasible paths. Ngo and Tan \cite{NgoT08a} proposed some execution trace patterns that witness likely infeasible paths. Delahaye \etal~\cite{DelahayeBG15} showed that infeasibility is caused by the same reason for many paths and thus, devised a technique that given an infeasible path can identify other, potentially unexplored paths. All these methods indirectly support test generation and contrary to ours do not detect polluting test objectives. 


\vspace{-1.5mm}
\subsection{Equivalent Mutants}
\label{subsec:related-equiv-mut}
\vspace{-1mm}

Automatically determining mutant equivalence is an instance of the infeasibility problem and is undecidable~\cite{OffuttP97}. There are numerous propositions on how to handle this problem, however most of them have only been evaluated on example programs and thus, their applicability and effectiveness remains unexplored \cite{7882714}. Due to space constraints we discuss the most recent and relevant approaches. Details regarding the older studies can be found in the recent paper by Kintis \etal \cite{7882714}, which extensively cover the topic.

One of the most recent methods is the Trivial Compiler Optimization (TCE) \cite{Papadakis15, 7882714}. The method assumes that equivalent mutant instances can be identified by comparing the object code of the mutants. The approach works well (it can identify 30\% of the equivalent mutants) as the compiler optimisations turn mutant equivalencies into the same object code. In contrast our approach uses state-of-the-art verification technologies (instead of  compilers) and targets all types of polluting objectives.

Alternative to static heuristics are the dynamic ones. Grun \etal \cite{GrunSZ09} and Schuler \etal \cite{SchulerDZ09} suggested measuring the impact of mutants on the program execution and program invariants in order to identify likely killable mutants. Schuler and Zeller \cite{SchulerZ13} investigate a large number of candidate impact measures and found that coverage was the most appropriate. Along the same lines Kintis \etal~\cite{KintisPM15} found that higher order mutants provide more accurate predictions than coverage.  
Overall, these approaches are unsound (they provide many false positives) and they  depend on the underlying test suites. In contrast our approach is sound and static.

\vspace{-1.5mm}
\subsection{Duplicate and Subsumed Test Objectives}
\label{subsec:related-equiv-mut}
\vspace{-1mm}

The problems caused by subsumed objectives have been identified a long time ago.
Chusho introduced  essential branches \cite{Chusho87},  
or non-dominated branches \cite{BertolinoM94}, as a way to prevent the inflation of the branch coverage score caused by redundant branches. He  also introduced a technique 
devising graph dominator analysis in order to identify the essential branches. Bertolino and Marr\'e \cite{BertolinoM94} also used  graph dominator analysis to reduce the number of test cases needed to cover test objectives and to help estimate the remaining testing cost. Although these approaches identify the harmful effects of redundant objectives, they rely on graph analysis, which results in a large number of false positives. Additionally, they cannot deal with infeasible objectives. 

In the context of mutation testing, Kintis \etal \cite{KintisPM10}  identified the problem and showed that mutant cost reduction techniques perform well when using all mutants but not when using non-redundant ones. Amman \etal \cite{AmmannDO14} introduced  minimal mutants and dynamic mutant subsumption and showed that mutation testing tools generate a large number of subsumed mutants. 

Although  mutant redundancies were known from the early days of mutation testing \cite{KurtzAODKG16},  their harmful effects were only recently realised. Papadakis \etal \cite{Papadakis16} performed a large-scale study and demonstrated that subsumed mutants inflate the mutation score measurement.
Overall, Papadakis \etal \cite{Papadakis16} showed that arbitrary experiments can result in different conclusions when they account for the cofounding effects of subsumed mutants. Similarly, Kurtz \etal \cite{KurtzAODKG16, KurtzAOK16} compared selective mutation testing strategies and found that they perform poorly when mutation score is free of redundant mutants.

Overall, most of the studies identify the problem but fail to deal with it. One attempt to reduce mutant redundancies uses TCE \cite{Papadakis15, 7882714} to remove duplicate mutants. Other attempts are due to Kurtz \etal \cite{KurtzAO15} who devised differential symbolic execution to identify subsumed mutants. Gong \etal \cite{GongZYM17} used dominator analysis (in the context of weak mutation) in order to reduce the number of mutants. Unfortunately, both studies have limited  scope as they have been evaluated only on example programs and their applicability and scalability remain unknown. Conversely, TCE is applicable and scalable, but it only targets specific kinds of subsumed mutants (duplicate ones) and cannot be applied on structural test objectives.

%% file: conclusion.tex
\vspace{-2mm}
\section{Conclusion}
\label{sec:conclusion}
\vspace{-1mm}

Software testing is the primary method for detecting software defects. 
In that context, polluting test objectives  are well-known to be harmful to the testing process, potentially wasting tester efforts and misleading them on the quality of their test suites.    
 We have presented LClean, the only approach to date that handles in a unified way the detection of (the three kinds of) polluting objectives  for a large set of common criteria,  
together with  a dedicated (open-source) tool able to prune out such polluting objectives.  LClean reduces the problem of detecting polluting objectives to the problem of proving assertions in the tested code. The tool relies on weakest-precondition calculus and SMT solving to prove these assertions. It is built on top of the industry-proof Frama-C  verification platform, 
 specifically  tuned to our  scalability needs. 
Experiments show that LClean provides  a useful, sound, scalable and adaptable means for helping testers to target high levels of coverage (where most faults are detected) and to evaluate more accurately the strength of their test suites (as well as of the tools possibly used to generate them).      
%
%
%
%
  A promising direction for future work is the extension of LClean  to the few remaining  unsupported classes of  test objectives, like data-flow  criteria.  %

%% file: appendices.tex
\appendix

\section*{Appendix: Detailed Results}

Figures \ref{fig-app1}, \ref{fig-app2}, \ref{fig-app3}, \ref{fig-app4}, \ref{fig-app5} and \ref{fig-app6} provide a more detailed view of the experimental results discussed in Sections \ref{sec:approach} and \ref{sec:evaluation}. 

{\footnotesize
\begin{figure*}[htbp]
\centering
\begin{tabular}{|c|c|c|c|c|c|}
\hline
{\textbf{Benchmark}} & {\textbf{\#Obj}}  & \textbf{{LClean}} & {\textbf{$\sum_{\textbf{Fun}}$(\#Obj/Fun)$^2$}}  & {\textbf{\#Obj$^2$}}  & {\textbf{Criterion}} \\
\hline
{\textbf{tcas}} & 56 & 152 & 976 & 3 136 & CC \\
\small{\textit{153 locs}} & 64 & 336 & 1 424 & 4 096 & MCC \\
& 56 & 152 & 976 & 3 136 & GACC \\
& 258 & 6 858 & 17 688 & 66 564 & WM \\
\hline
{\textbf{schedule2}} & 76 & 148 & 672 & 5 776 & CC \\
\small{\textit{307 locs}} & 76 & 148 & 672 & 5 776 & MCC \\
& 76 & 148 & 672 & 5 776 & GACC \\
& 331 & 3 362 & 15 529 & 109 561 & WM \\
\hline
{\textbf{schedule}} & 58 & 114 & 460 & 3 364 & CC \\
\small{\textit{410 locs}} & 58 & 114 & 460 & 3 364 & MCC \\
&58 & 114 & 460 & 3 364& GACC \\
& 253 & 2 478 & 9 343 & 64 009 & WM \\
\hline
{\textbf{totinfo}} & 88 & 144 & 2 144 & 7 744 & CC \\
\small{\textit{406 locs}} & 88 & 144 & 2 144 & 7 744  & MCC \\
&  88 & 144 & 2 144 & 7 744 & GACC \\
& 884 & 26 250 & 177 618 & 781 456 & WM \\
\hline
{\textbf{printokens2}} & 162 & 450 & 2 564 & 26 244 & CC \\
\small{\textit{499 locs}} & 164 & 476 & 2 640 & 26 896 & MCC \\
& 162 & 450 & 2 564 & 26 244 & GACC \\
& 526 & 7 098 & 37 554 & 276 676 & WM \\
\hline
{\textbf{printtokens}} & 66 & 74 & 444 & 4 356 & CC \\
\small{\textit{794 locs}} & 66 & 74 & 444 & 4 356 & MCC \\
& 66 & 74 & 444 & 4 356 & GACC \\
& 362 & 2 654 & 15 516 & 131 044 & WM \\
\hline
{\textbf{replace}} & 148 & 312 & 2 624 & 21 904 & CC \\
\small{\textit{641 locs}} & 150 & 342 & 2 756 & 22 500 & MCC \\
& 148 & 312 & 2 624 & 21 904 & GACC \\
& 929 & 17 934 & 73 711 & 863 041 & WM \\
\hline
{\textbf{err}} & 202 & 714 & 3 228 & 40 804 & CC \\
\small{\textit{1201 locs}} & 202 & 714 & 3 228 & 40 804 & MCC \\
& 202 & 714 & 3 228 & 40 804 & GACC \\
& 768 & 12 168 & 39 812 & 589 824 & WM \\
\hline
{\textbf{aes\_core}} & 36 & 76 & 120 & 1296 & CC \\
\small{\textit{1375 locs}} & 38 & 102 & 828 & 1444 & MCC \\
& 36 & 76 & 120 & 1296  & GACC \\
& 883 & 54 432 & 209 231 & 779 689 & WM \\
\hline
{\textbf{pem\_lib}} & 304 & 680 & 13 416 & 92 416 & CC \\
\small{\textit{904 locs}} & 330 & 1 634 & 16 004 & 108 900 & MCC \\
&  304 & 680 & 13 416 & 92 416 & GACC \\
& 1 149 & 13 448 & 206 467 & 1 320 201 & WM \\
\hline
{\textbf{bn\_exp}} & 480 & 1 144 & 37 360 & 230 400 & CC \\
\small{\textit{1498 locs}} & 596 & 5 732 & 59 232 & 355 216 & MCC \\
&  480 & 1 144 & 37 360 & 230 400 & GACC \\
& 2 182 & 35 200 & 687 808 & 4 761 124 & WM \\
\hline
{\textbf{gzip}} & 1670 & 6 166 & 71 580 & 2 788 900 & CC \\
\small{\textit{7569 locs}} & 1 726 & 7 558 & 79 268 & 2 979 076 & MCC \\
& 1670 & 6 166 & 71 580 & 2 788 900 & GACC \\
& 12 270 & 323 244 & 3 967 728 & 150 552 900 & WM \\
\hline
{\textbf{sjeng chess ia}} & 4 090 & 20 290 & 696 156 & 16 728 100 & CC \\
\small{\textit{14070 locs}} & 4 746 & 60 306 & 1 002 956 & 22 524 516 & MCC \\
& 4 090 & 20 290 & 696 156 & 16 728 100 & GACC \\
& 25 722 & 644 754 & 25 101 370 & 661 621 284 & WM \\
\hline
{\textbf{sqlite}} & 20 202 & 63 578 & 2 182 196 & 408 120 804 & CC \\
\small{\textit{196888 locs}} & 21 858 & 236 594 & 2 788 948 & 477 772 164 & MCC \\
& 20 202 & 63 578 & 2 182 196 & 408 120 804 & GACC \\
& 90 410 & 1 761 028 & 49 603 128 & 8 173 968 100 & WM \\
\hline
\hline
{\textbf{TOTAL}} & 27 638 & 94 042 & 3 013 940 & 428 075 244 & MCC \\
\small{\textit{226333 locs}} & 30 162 & 314 274 & 3 961 004 & 503 856 852 & GACC \\
& 27 638 & 94 042 & 3 013 940 & 428 075 244 & WM \\
& 136 927 & 2 910 908 & 80 162 503 & 8 995 885 473 & CC \\
\cline{2-5}
& 222 365 & 3 413 266 & 90 151 387 $\mathbf{\color{red}(\times 26)}$ & 10 355 892 813 $\mathbf{\color{red}(\times 3034)}$ & all \\
\hline
\end{tabular} 
\caption{Number of pairs of labels in 14 C programs (detailed version of Figure \ref{fig-n})}
\label{fig-app1}
\end{figure*}}

{\footnotesize
\begin{figure*}[htbp]
\centering
\begin{tabular}{|c|c|c|c|c|c|c|c|c|c|c|c|c|}
\hline 
\multirow{3}{*}{\textbf{Benchmark}} & \multirow{3}{*}{\textbf{\#Objectives}} & \multicolumn{2}{c|}{\textbf{STEP 1}}  & \multicolumn{2}{c|}{\textbf{STEP 2}} & \multicolumn{3}{c|}{\textbf{STEP 3}}  & \multicolumn{3}{c|}{\textbf{TOTAL}} & \multirow{3}{*}{\textbf{Criterion}} \\
 & & \textit{pruned} & \multirow{2}{*}{\textit{time}} & \textit{pruned} & \multirow{2}{*}{\textit{time}} & \textit{pruned}  & \textit{pruned} & \multirow{2}{*}{\textit{time}} & \multicolumn{2}{c|}{\textit{pruned}} & \multirow{2}{*}{\textit{time}} &  \\ 
  & & \textit{infeasible} &  & \textit{duplicate} & & \textit{duplicate}  & \textit{subsumed} & & \textit{ratio} & \textit{\%} & &  \\ 
\hline 
{\textbf{tcas}}  & 56 & 0 & 2s & 0 & 3s & 2 & 4  
& 9s & 6/56 & 11\% & 14s & CC \\
\scriptsize {\textit{153 locs}} & 64 & 8 & 3s & 0 & 4s & 0 & 0 & 7s    
& 8/64 & 13\% & 14s & MCC \\ 
 & 56 & 1 & 3s & 0 & 3s & 1 & 0  
& 8s & 2/56 & 4\% & 14s & GACC \\ 
 & 258 & 8 & 11s & 13 & 10s & 9 & 31  
& 45s &  61/258 & 24\% & 66s & WM \\  
\hline 
{\textbf{schedule2}}  & 76 & 0 & 3s & 0 & 4s & 0 & 8   
& 6s & 8/76 & 11\% & 13s & CC \\
\scriptsize {\textit{307 locs}} & 76 & 4 & 3s & 0 & 4s & 0 & 0    
& 5s & 4/76 & 5\% & 12s & MCC \\ 
 & 76 & 0 & 3s & 0 & 4s & 5 & 1  
& 6s & 6/76 & 8\% & 13s & GACC \\ 
 & 331 & 4 & 8s & 15 & 9s & 21 & 29  
& 29s &  69/331 & 21\% & 46s & WM \\ 
\hline 
{\textbf{schedule}}  & 58 & 0 & 3s & 0 & 3s & 0 & 6  
& 8s & 6/58 & 10\% & 14s & CC \\
\scriptsize {\textit{410 locs}} & 58 & 3 & 3s & 0 & 3s & 0 & 0   
& 4s & 3/58 & 5\% & 10s & MCC \\ 
 & 58 & 0 & 4s & 0 & 4s & 2 & 0  
& 5s & 2/58 & 3\% & 13s & GACC \\ 
 & 253 & 2 & 6s & 6 & 7s & 12 & 19 
& 23s &  39/253 & 15\% & 36s & WM \\ 
\hline 
{\textbf{totinfo}}  & 88 & 0 & 11s & 0 & 11s & 0 & 0  
& 13s & 0/88 & 0\% & 35s & CC \\
\scriptsize {\textit{406 locs}} & 88 & 0 & 11s & 0 & 11s & 0 & 0    
& 13s & 0/88 & 0\% & 35s & MCC \\ 
 & 88 & 0 & 11s & 0 & 11s & 1 & 0  
& 13s & 1/88 & 3\% & 35s & GACC \\ 
 & 884 & 0 & 35s & 12 & 35s & 16 & 47  
& 313s &  75/884 & 8\% & 383s  & WM \\ 
\hline 
{\textbf{printtokens2}}  & 162 & 0 & 4s & 0 & 4s & 0 & 15   
& 27s & 15/162 & 9\% & 35s & CC \\
\scriptsize {\textit{499 locs}} & 164 & 5 & 4s & 0 & 4s & 0 & 10    
& 27s & 15/164 & 9\% & 35s & MCC \\ 
 & 162 & 0 & 4s & 0 & 4s & 2 & 10  
& 27s & 12/162 & 7\% & 35s & GACC \\ 
 & 526 & 3 & 9s & 19 & 9s & 22 & 87
& 70s & 131/526 & 25\% & 88s  & WM \\ 
\hline 
{\textbf{printtokens}}  & 66 & 0 & 6s & 0 & 7s & 0 & 2   
& 7s & 2/66 & 3\% & 20s & CC \\
\scriptsize {\textit{794 locs}} & 66 & 0 & 6s & 0 & 7s & 0 & 2   
& 7s & 2/66 & 3\% & 20s & MCC \\ 
 & 66 & 0 & 6s & 0 & 7s & 0 & 2  
& 7s & 2/66 & 3\% & 20s & GACC \\ 
 & 362 & 3 & 18s & 24  & 14s & 13 & 56  
& 48s & 96/362 & 27\% & 80s & WM \\ 
\hline 
{\textbf{replace}}  & 148 & 0 & 6s & 0 & 6s & 0 & 6  
& 13s & 6/148 & 4\% & 25s & CC \\
\scriptsize {\textit{641 locs}} & 150 & 0 & 6s & 0 & 7s & 0 & 4   
& 15s & 4/150 & 3\% & 28s & MCC \\ 
 & 148 & 0 & 6s & 0 & 6s & 7 & 4 
& 11s & 11/148 & 7\% & 23s & GACC \\ 
 & 929 & 17 & 27s & 34 & 42s & 41 & 67  
& 195s & 159/929 & 17\% & 264s & WM \\ 
\hline
\hline  
{\textbf{TOTAL}}  & 654 & 0 & 35s & 0 & 38s & 2 & 41    
& 83s & 43/654 & \textbf{7\%} & \textbf{156s} & \textbf{CC} \\
\scriptsize {\textit{3210 locs}} & 666 & 20 & 36s & 0 & 40s & 0 & 16    
& 78s & 36/666  &  \textbf{5\%} & \textbf{154s}  & \textbf{MCC} \\ 
 & 654 & 1 & 37s & 0 & 39s & 18 &   
 17 & 77s & 36/654 &  \textbf{6\%} & \textbf{153s}  & \textbf{GACC} \\ 
 & 3543 & 37 & 114s & 123 & 126s & 134 & 336 
& 723s & 630/3543  &  \textbf{18\%} & \textbf{963s}   & \textbf{WM} \\ 
\cline{2-13} 
 & 5517 & 58 & 222s & 123 & 243s & 154 & 410  
& 961s & 745/5517  &  \textbf{\color{red}14\%} & \textbf{\color{red}23m46s}   & \textbf{\color{red}All} \\ 
\hline 
\end{tabular} 
\caption{Pruning power and computation time of LClean over the Siemens programs (details from Figure \ref{fig-all})}
\label{fig-app2}
\end{figure*}}

{\footnotesize 
\begin{figure*}[htbp]
\centering
\begin{tabular}{|c|c|c|c|c|c|c|c|c|c|}
\hline 
\multirow{2}{*}{\textbf{Benchmark}} & \multirow{2}{*}{\textbf{\#Objectives}} & \multicolumn{2}{c|}{\textbf{STEP 1}}  & \multicolumn{2}{c|}{\textbf{STEP 2}} & \multicolumn{2}{c|}{\textbf{STEP 3}}  & \multicolumn{2}{c|}{\textbf{TOTAL}} \\
 & & \textit{no optim} & {\textit{multi-core}} & \textit{no optim} & {\textit{multi-core}} & \textit{no optim} & {\textit{multi-core}}  & \textit{no optim} & {\textit{multi-core}}   \\ 
\hline 
{\textbf{tcas}} 
 & 258 & 368s & 20s &  368s  & 21s  
& 2925s & 149s & 3661s & 190s  \\  
\hline 
{\textbf{schedule2}}  
 & 331  & 387s & 21s & 420s  & 22s   
& 1626s & 84s & 2433s  & 127s \\   
\hline 
{\textbf{schedule}} 
 & 253  & 292s & 16s & 295s & 16s   
& 1222s & 63s & 1809s  & 95s \\   
\hline 
{\textbf{totinfo}}  
 & 884  & 1568s & 93s & 1646s & 99s   
& 14182s & 736s & 17396s  & 928s  \\  
\hline 
{\textbf{printtokens2}}
 & 526 & 572s & 37s & 553s & 37s    
& 3567s & 182s & 4692s  & 256s  \\   
\hline 
{\textbf{printtokens}}  
 & 362  & 430s & 32s & 411s & 30s    
& 1571s & 89s & 2412s & 151s \\  
\hline 
{\textbf{replace}}  
 & 929  & 1431s & 83s & 1403s & 93s    
& 8453s & 428s & 11287s & 604s  \\  
\hline
\hline  
{\textbf{TOTAL}}  
 & 3543  & 5048s  $\mathbf{\color{red}(\times 44)}$  & 302s  $\mathbf{\color{red}(\times 2.6)}$ & 5096s  $\mathbf{\color{red}(\times 40)}$  & 318s  $\mathbf{\color{red}(\times 2.5)}$   
& 33546s $\mathbf{\color{red}(\times 46)}$  & 1731s $\mathbf{\color{red}(\times 2.4)}$ & 43690s $\mathbf{\color{red}(\times 45)}$ & 2351s $\mathbf{\color{red}(\times 2.4)}$ \\   
\hline 
\end{tabular} 
\caption{Tool optimizations impact (Siemens, WM) (detailed version of Figure \ref{fig-impact-optims}))}
\label{fig-app3}
\end{figure*}}

{\footnotesize
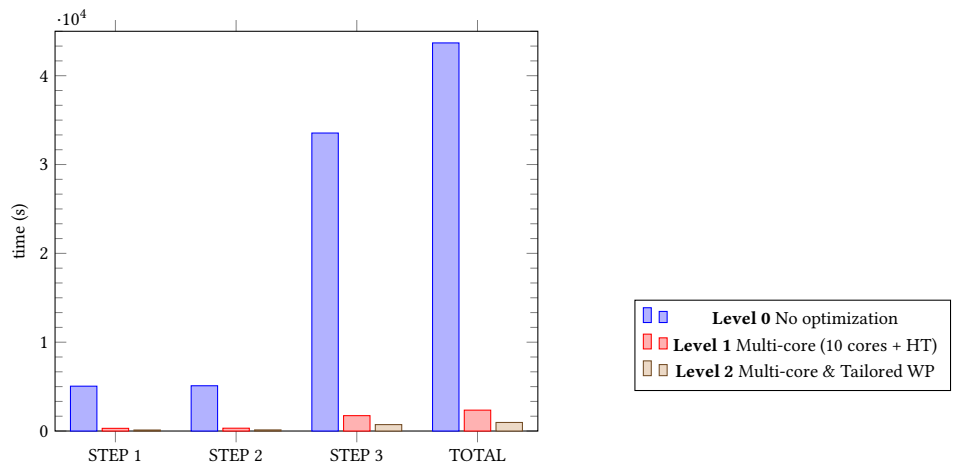
\begin{figure*}[htbp]
\begin{center}
\pgfplotsset{every axis legend/.append style={
at={(1.2,0.1)},
anchor=south west}}
\noindent\begin{tikzpicture}[baseline]
\begin{axis}[width=8cm,ybar=2pt,symbolic x coords={x,STEP 1,y,STEP 2,z,STEP 3,a,TOTAL,b}, xtick=data,
xmin=x,
xmax=b,
ylabel={time (s)},
y label style={at={(axis description  cs:0.15,.5)},anchor=south},
ymin=0,
ymax=45000,
minor y tick num=5,
legend entries={\textbf{Level 0} No optimization,\textbf{Level 1} Multi-core (10 cores + HT),\textbf{Level 2} Multi-core \& Tailored WP},
]
\addplot coordinates {(STEP 1,5048) (STEP 2,5096) (STEP 3,33546.0) (TOTAL,43690.0)};  
\addplot coordinates {(STEP 1,302) (STEP 2,318) (STEP 3,1731.0) (TOTAL,2351.0)}; 
\addplot coordinates {(STEP 1,114) (STEP 2,126) (STEP 3,723) (TOTAL,963)};  
\end{axis} \end{tikzpicture}
\end{center}
\caption{Tool optimizations impact (Siemens, WM) (non-logarithmic version of Figure \ref{fig-impact-optims})}
\label{fig-app4}
\end{figure*}}

{\footnotesize
\begin{figure*}[htbp]
\centering
\begin{tabular}{|c|c|c|c|c|cLL|c|c|}
\hline 
\multirow{2}{*}{\textbf{Benchmark}} & \multirow{2}{*}{\textbf{\#Objectives}} & \multicolumn{4}{c}{\textbf{\#Prunable polluting objectives}} & 
& & \multirow{2}{*}{\textbf{Time}} & \multirow{2}{*}{\textbf{Criterion}} \\
 & & \textit{infeasible} & \textit{duplicate} & \textit{subsumed} & \textit{total} & \textit{raw} & \textit{LClean} &  &  \\
\hline 
{\textbf{tcas}} & 56 & 4 & 23 & 15 & 42 & 52/56 (93\%) & 46/50 (92\%) & 188s & CC \\
\scriptsize {\textit{153 locs}}& 64 & 13 & 21 & 13 & 47 & 51/64 (80\%) & 51/56 (91\%) & 198s & MCC \\
& 56 & 7 & 22 & 14 & 43 & 49/56 (87\%) & 49/54 (91\%) & 194s & GACC \\
& 258 & 19 & 130 & 81 & 230 & 239/258 (93\%) & 186/197 (94\%) & 329s & WM \\
\hline 
{\textbf{schedule2}} & 76 & 5 & 24 & 28 & 57 & 71/76 (93\%) & 63/68 (93\%) & 503s & CC \\
\scriptsize {\textit{307 locs}}& 76 & 12 & 21 & 24 & 57 & 64/76 (84\%) & 64/72 (89\%) & 501s & MCC \\
& 76 & 5 & 28 & 24 & 57 & 71/76 (93\%) & 65/70 (93\%) & 509s & GACC \\
& 331 & 33 & 207 & 61 & 301 & 298/331 (90\%) & 238/262 (91\%) & 1323s & WM \\
\hline 
{\textbf{schedule}} & 58 & 3 & 7 & 26 & 36 & 55/58 (95\%) & 49/52 (94\%) & 400s & CC \\
\scriptsize {\textit{410 locs}}& 58 & 7 & 5 & 21 & 33 & 51/58 (88\%) & 51/55 (93\%) & 403s & MCC \\
& 58 & 3 & 10 & 21 & 34 & 55/58 (95\%) & 53/56 (95\%) & 399s & GACC \\
& 253 & 13 & 151 & 59 & 223 & 240/253 (95\%) & 203/214 (95\%) & 810s & WM \\
\hline 
{\textbf{totinfo}} & 88 & 9 & 34 & 29 & 72 & 79/88 (90\%) & 79/88 (90\%) & 204s & CC \\
\scriptsize {\textit{406 locs}}& 88 & 9 & 35 & 27 & 71 & 79/88 (90\%) & 79/88 (90\%) & 197s & MCC \\
& 88 & 9 & 36 & 27 & 72 & 79/88 (90\%) & 78/87 (90\%) & 197s & GACC \\
& 884 & 74 & 679 & 104 & 857 & 810/884 (92\%) & 741/809 (92\%) & 1090s & WM \\
\hline 
{\textbf{printtokens2}} & 162 & 7 & 80 & 40 & 127 & 155/162 (96\%) & 140/147 (95\%) & 864s & CC \\
\scriptsize {\textit{499 locs}}& 164 & 13 & 78 & 38 & 129 & 151/164 (92\%) & 141/149 (95\%) & 859s & MCC \\
& 162 & 8 & 82 & 37 & 127 & 154/162 (95\%) & 142/150 (95\%) & 855s & GACC \\
& 526 & 71 & 349 & 80 & 500 & 455/526 (87\%) & 336/395 (85\%) & 2085s & WM \\
\hline 
{\textbf{printtokens}} & 66 & 5 & 29 & 20 & 54 & 61/66 (92\%) & 59/64 (92\%) & 694s & CC \\
\scriptsize {\textit{794 locs}}& 66 & 5 & 29 & 20 & 54 & 61/66 (92\%) & 59/64 (92\%) & 694s & MCC \\
& 66 &5 & 29 & 20 & 54 & 61/66 (92\%) & 59/64 (92\%) & 694s & GACC \\
& 362 & 41 & 257 & 51 & 349 & 321/362 (89\%) & 230/266 (86\%) & 1599s & WM \\
\hline 
{\textbf{replace}} & 148 & 4 & 46 & 72 & 122 & 144/148 (97\%) & 138/142 (97\%) & 279s & CC \\ 
\scriptsize {\textit{641 locs}}& 150 & 17 & 32 & 72 & 121 & 133/150 (89\%) & 129/146 (88\%) & 290s & MCC \\
& 148 & 9 & 42 & 69 & 120 & 139/148 (94\%) & 128/137 (93\%) & 286s & GACC \\
& 929 & 135 & 554 & 205 & 894 & 794/929 (85\%) & 661/770 (86\%) & 1163s & WM \\
\hline
\hline  
{\textbf{TOTAL}} & 654 & 37 & 243 & 230 & 510 &	617/654	(94\%)	&	574/611	(94\%)	& 3132s & CC \\
\scriptsize {\textit{3210 locs}} & 666 & 76 & 221	 & 215 & 512 &	590/666	(89\%)	&	574/630	(91\%)	&3142s & MCC \\
& 654 & 46 & 249 & 212 & 507 &	608/654	(93\%)	&	574/618	(93\%)	& 3134s & GACC \\
& 3543 & 386 & 2327 & 641 & 3354 &	3157	/3543 (89\%)	&	2595/2913 (89\%)	& 8399s & WM \\
\cline{2-10}
& 5517 & 545 & 3040 & 1298 & 4883 $\mathbf{\color{red}(89\%)}$ &	4972	/5517 (90\%)	&	4317/4772 (90\%)	& 
4h56m47 $\mathbf{\color{red}(\times 12)}$
& All \\
\hline 
\end{tabular} 
\caption{Dynamic detection of (likely) polluting objectives (details from Figure \ref{fig-dynamic})}
\label{fig-app5}
\end{figure*}}

{\footnotesize
\begin{figure*}[htbp]
\centering
\begin{tabular}{|c|c|c|c|c|c|c|c|c|}
\hline 
\multirow{2}{*}{\textbf{Benchmark}} & \multirow{2}{*}{\textbf{\#Objectives}} & \multicolumn{2}{c|}{\textbf{LUncov}} & \multicolumn{2}{c|}{\textbf{LClean} (infeasible only)} & \multicolumn{2}{c|}{\textbf{LClean} (all polluting)}  &
  \multirow{2}{*}{\textbf{Criterion}}\\
   &  & \textit{pruned} & \textit{time} & \textit{pruned} & \textit{time} & \textit{pruned} & \textit{time} & \\ 
\hline
{\textbf{trityp}} & 24 & 0 & 13s & 0 & 1s & 0 & 5s & CC \\
\scriptsize {\textit{50 locs}} & 28 & 0 & 14s & 0 & 2s & 0 & 7s & MCC \\
& 129 & 4 & 75s & 4 & 4s & 40 & 30s & WM \\
\hline
{\textbf{fourballs}} \scriptsize {\textit{35 locs}} & 67 & 11 & 26s & 11 & 2s & 26 & 11s & WM \\
\hline
{\textbf{utf8-3}} \scriptsize {\textit{108 locs}} & 84 & 29 & 140s & 2 & 14s & 53 & 101s & WM \\
\hline
{\textbf{utf8-5}} \scriptsize {\textit{108 locs}} & 84 & 2 & 140s & 2 & 15s & 51 & 99s & WM \\
\hline
{\textbf{utf8-7}} \scriptsize {\textit{108 locs}} & 84 & 2 & 140s & 2 & 15s & 54 & 102s & WM \\
\hline
{\textbf{tcas}} & 10 & 0 & 5s & 0 & 1s & 0 & 4s & CC \\
\scriptsize {\textit{124 locs}} & 12 & 1 & 6s & 1 & 1s & 1 & 4s & MCC \\
& 111 & 10 & 61s & 6 & 7s & 18 & 35s & WM \\
\hline
{\textbf{replace}} \scriptsize {\textit{100 locs}} & 80 & 10 & 40s & 3 & 2s & 43 & 13s & WM \\
\hline
{\textbf{full\_bad}} & 16 & 4 & 9s & 4 & 2s & 7 & 5s & CC \\
\scriptsize {\textit{219 locs}} & 39 & 15 & 21s & 15 & 2s & 20 & 6s & MCC \\
& 46 & 11 & 25s & 9 & 2s & 16 & 10s & WM \\
\hline
{\textbf{get\_tag-5}} & 20 & 0 & 11s & 0 & 2s & 8 & 6s & CC \\
\scriptsize {\textit{240 locs}} & 26 & 0 & 15s & 0 & 2s & 8 & 9s & MCC \\
& 47 & 2 & 28s & 0 & 2s & 13 & 11s & WM \\
\hline
{\textbf{get\_tag-6}} & 20 & 0 & 11s & 0 & 2s & 8 & 8s & CC \\
\scriptsize {\textit{240 locs}} & 26 & 0 & 15s & 0 & 2s & 8 & 9s & MCC \\
& 47 & 2 & 28s & 0 & 2s & 13 & 12s & WM \\
\hline
{\textbf{gd-5}} & 36 & 0 & 23s & 0 & 2s & 14 & 9s & CC \\
\scriptsize {\textit{319 locs}} & 36 & 7 & 27s & 7 & 3s & 7 & 9s & MCC \\
& 63 & 1 & 49s & 0 & 5s & 29 & 19s & WM \\
\hline
{\textbf{gd-6}} & 36 & 0 & 25s & 0 & 2s & 14 & 9s & CC \\
\scriptsize {\textit{319 locs}} & 36 & 7 & 27s & 7 & 3s & 7 & 9s & MCC \\
& 63 & 0 & 49s & 0 & 5s & 29 & 20s & WM \\
\hline
\hline
{\textbf{TOTAL}} & 162 & 4 & 97s & 4 & 12s & 51 & 46s & CC \\
\scriptsize {\textit{1970 locs}} & 203 & 30 & 125s & 30 & 15s & 51 & 53s & MCC \\
& 905 & 84 & 801s & 41 & 75s & 385 & 463s & WM \\
\cline{2-9}
& \textbf{1270} & \textbf{9\%} 
& \textbf{17m3s} 
&  \textbf{6\%} $\mathbf{\color{red}(\div 1.6)}$ 
& \textbf{1m42s} $\mathbf{\color{blue}(\div 10)}$ 
& \textbf{38\%} $\mathbf{\color{red}(\times 4.2)}$  
&\textbf{9m22s} $\mathbf{\color{blue}(\div 1.8)}$ 
& \textbf{all} \\
\hline
\end{tabular} 
\caption{LUncov \cite{bardin15} vs LClean (benchmarks from \cite{bardin15}) (detailed version of Figure \ref{fig-luncov})}
\label{fig-app6}
\end{figure*}}